# Lost in Instructions: Study of Blind Users' Experiences with DIY Manuals and AI-Rewritten Instructions for Assembly, Operation, and Troubleshooting of Tangible Products

Monalika Padma Reddy
Department of Computer Science
Stony Brook University
Stony Brook, New York, USA
mpadmareddy@cs.stonybrook.edu

Aruna Balasubramanian
Computer Science
Stony Brook University
Stony Brook, New York, USA
arunab@cs.stonybrook.edu

Jiawei Zhou
Applied Math and Statistics
Stony Brook University
Stony Brook, New York, USA
jiawei.zhou.1@stonybrook.edu

Xiaojun Bi
Department of Computer Science
Stony Brook University
Stony Brook, New York, USA
xiaojun@cs.stonybrook.edu

IV Ramakrishnan
Computer Science
Stony Brook University
Stony Brook, New York, USA
ram@cs.stonybrook.edu

Vikas Ashok
Department of Computer Science
Old Dominion University
Norfolk, Virginia, USA
vganjigu@odu.edu

## Abstract

AI tools like ChatGPT and Be-My-AI are increasingly being used by blind individuals. Although prior work has explored their use in some Do-It-Yourself (DIY) tasks by blind individuals, little is known about how they use these tools and the available product-manual resources to assemble, operate, and troubleshoot physical/tangible products – tasks requiring spatial reasoning, structural understanding, and precise execution. We address this knowledge gap via an interview study and a usability study with blind participants, investigating how they leverage AI tools and product manuals for DIY tasks with physical products. Findings show that manuals are essential resources, but product-manual instructions are often inadequate for blind users. AI tools presently do not adequately address this insufficiency, in fact, we observed that they often exacerbate this issue with incomplete, incoherent, or misleading guidance. Lastly, we suggest improvements to AI tools for generating tailored instructions for blind users' DIY tasks involving tangible products.

**ACM Reference Format:**
Monalika Padma Reddy, Aruna Balasubramanian, Jiawei Zhou, Xiaojun Bi, IV Ramakrishnan, and Vikas Ashok. 2026. Lost in Instructions: Study of Blind Users' Experiences with DIY Manuals and AI-Rewritten Instructions for Assembly, Operation, and Troubleshooting of Tangible Products. In *Proceedings of the 2026 CHI Conference on Human Factors in Computing Systems (CHI '26), April 13–17, 2026, Barcelona, Spain.* ACM, New York, NY, USA, 28 pages. https://doi.org/10.1145/3772318.3790955

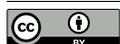

## 1 Introduction

Blind individuals encounter persistent difficulties in accessing and interpreting instructional materials, primarily due to the visual nature of the content and the presence of many interrelated components that are difficult to understand without accessible, well-structured alternatives within these support resources [7, 21, 35, 61, 101, 110]. Over the past two decades, assistive technologies have sought to address these barriers. Tools such as KNFB Reader [98] and TapTapSee [1] use computer vision techniques such as optical character recognition (OCR) and object detection and recognition to access printed text, identify objects, and generate scene-level descriptions. While effective for basic interpretation, they lack support for follow-up queries and do not provide more nuanced or detailed understanding.

The increased availability of Large Language Models (LLMs) and Vision-Language Models (VLMs), such as GPT-4 [3], signifies a paradigmatic shift in assistive technologies for blind users. AI-powered assistive tools such as Be-My-AI[1] and general-purpose AI tools like ChatGPT[2] enable blind users to pose complex queries, obtain contextually-rich explanations, and engage with visual information in more flexible and interactive ways. In doing so, these tools reduce reliance on sighted assistance and promote independence in everyday tasks such as shopping and navigating unfamiliar spaces, quickly becoming indispensable resources for blind users. This widespread adoption has spurred an extensive line of research investigating how such AI tools support accessible information access, knowledge acquisition, content creation, and daily activities such as social interactions and style-related decisions, while also identifying their limitations [6, 32, 51, 96, 120, 133].

While extant works on AI tools[3] have covered a wide variety of scenarios including scene description, navigation, and social interactions [51, 55, 96, 133] as well as many household DIY activities such as cooking and laundry [65, 133], little is known about how blind users leverage available support manuals [4] and AI tools for DIY tasks involving household products, specifically assembling components, learning functionalities, and troubleshooting problems. Such DIY tasks demand accurate interpretation of product-manual content, correct product-component identification, clear spatial reasoning and structural understanding, and precise execution of step-by-step directions. The margin for error is especially narrow; even minor misalignments at early stages can hinder later progress or even compromise overall product functionality. Therefore, the onus is high on AI tools to provide operationalizable non-visual stepwise instructions that blind users can understand and act on, by accurately translating the product-manual instructions that are predominantly vision-centric [19, 35]. By examining how blind users engage with product manuals and AI tools in the context of product-related DIY tasks, this paper investigates existing support gaps in AI-driven instructional assistance and highlights the unique challenges these tasks pose in practice. To guide this investigation, we pose the following research questions:

- **RQ1:** To what extent do blind users rely on product manuals, and what assistive tools do they employ to access and interpret the instructional content in these manuals?
- **RQ2:** What challenges do blind users face when using product manuals and AI tools for DIY tasks, and what workaround strategies do they employ to overcome these challenges?
- **RQ3:** What are the needs and preferences of blind users when engaging with product manuals and AI tools for product-related DIY tasks?

---
[1] https://www.bemyeyes.com/bme-ai/
[2] https://chatgpt.com/
[3] By "AI tools", we refer to both applications designed for blind users, e.g., *Be-My-AI, Seeing-AI* and general-purpose AI tools like *ChatGPT*.
[4] We use the terms support manuals, product manuals and OEM manuals interchangeably.

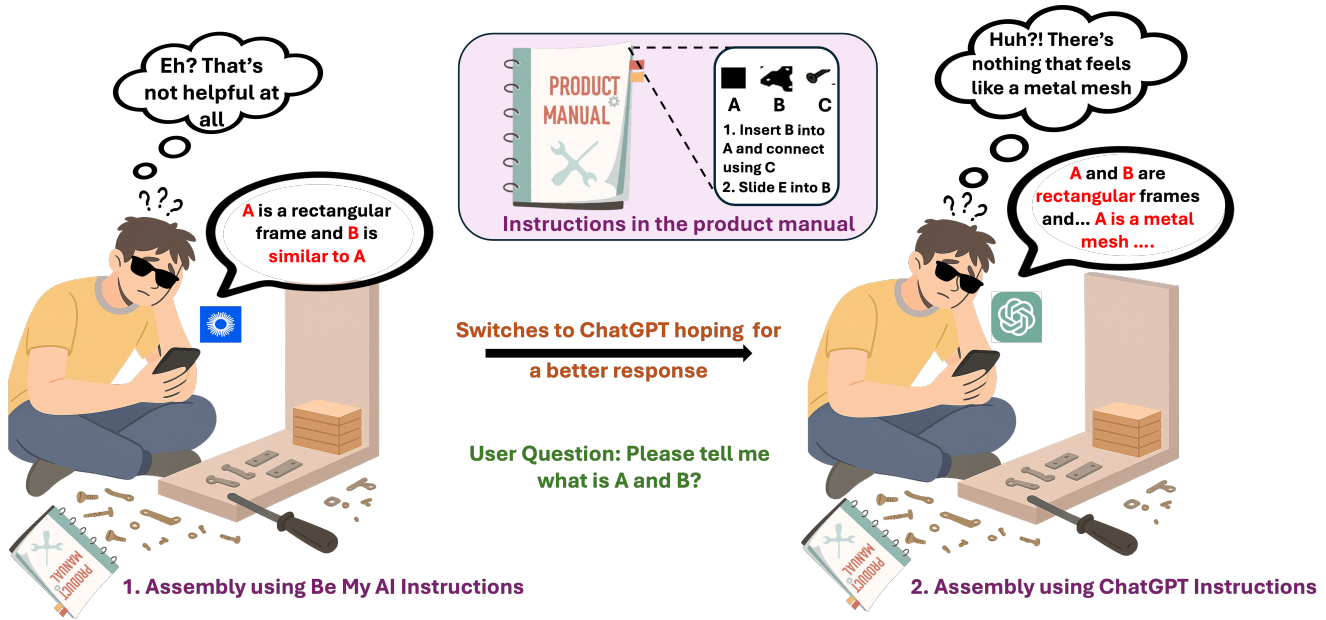

**Figure 1: Illustration of support gaps in AI-assistive tools for product-related DIY tasks, in this case, interpreting an assembly manual after a blind user captures an image of it. In both cases, the AI tools rely on inaccessible visual referents and misidentify component B: (Left) Be-My-AI gives only a vague incorrect reference (e.g., "B is similar to A"), while (Right) ChatGPT incorrectly describes B as a rectangular frame and further introduces a fictitious feature by calling A "a metal mesh." Both cases illustrate how current AI tools either under-specify or hallucinate, forcing blind users into dead ends during DIY product assembly.**

To answer these questions, we conducted two complementary studies with blind participants: a semi-structured interview study and a usability study. First, we interviewed 15 blind participants about their lived experiences with AI tools and support manuals while performing product-related DIY tasks, capturing both their present challenges with these resources and their needs/preferences for how instructional information should be conveyed for effective task execution. Second, we conducted a usability study to observe blind users performing DIY tasks with AI support in real time, enabling us to capture behavioral patterns, nuanced challenges, and failure patterns that impeded independent task completion, which interviews alone could not reveal.

Our study revealed that blind users relied primarily on traditional product manuals as the *most authoritative source of information* for DIY tasks, although the extent of reliance varied with task type and the presentation modality of instructional content. Participants consistently turned to AI tools to transform vision-centric manual instructions into non-visual, operationalizable, stepwise guidance. However, these AI-generated instructions frequently fell short (See Fig. 1), severely reducing efficiency and accuracy: none of the participants were able to complete tasks independently, and overall step-level accuracy remained below 50%. Participants also provided concrete suggestions for transforming product manual instructions into formats better suited to the needs of blind users. Drawing on these insights, we propose design directions for AI systems capable of generating context-sensitive, accessible guidance to support blind users in completing product-related DIY tasks more effectively and independently.

Overall, our paper makes two key contributions. First, we present an in-depth empirical account of how blind users leverage AI tools such as Be-My-AI and ChatGPT along with traditional resources like support manuals, for product-related DIY tasks, extending prior research on the use of AI systems to support autonomy among blind users. Second, we identify the support gaps in existing AI tools for product-manual-mediated DIY tasks, and analyze how challenges including vision-default bias, hallucinations, further exacerbate these support gaps by undermining both efficiency and accuracy.

## 2 Background and Related Work

Our study is situated within three strands of related work pertaining to: (i) the use of AI tools for accessing visual information; (ii) the challenges of blind individuals in Instrumental Activities of Daily Living (IADLs); and (iii) extant methods supporting non-visual access to printed and PDF manuals.

### 2.1 AI-Assistive Tools for Blind Users

Blind users rely on assistive technologies to access visual information by translating it into natural language descriptions [11]. Based on how the information is conveyed, they are classified into two categories: human-assisted or AI-driven. Human-assisted services connect users with trained agents (e.g., Aira[5]), volunteers [18], or crowd workers [22] who provide real-time descriptions in response to user requests. Although these systems provide accurate and nuanced descriptions, timely access depends on operator availability [16, 78, 79].

Consequently, blind users increasingly rely on AI-driven applications. Assistive applications such as TapTapSee [1], and KNFB Reader [98] support everyday activities like navigation [32, 51], and shopping by leveraging computer vision techniques such as OCR and object recognition. Recent advances in VLMs extend these capabilities further. AI-assistive tools such as Be-My-AI, powered by GPT-4 [3], enables richer scene understanding, interactive querying, and real-time feedback for capturing higher-quality images [6, 52, 133]. Recently, researchers have gone beyond commercially available solutions by developing assistive systems that adapt LLMs and VLMs to help blind users interact with GUIs [73, 95, 105], generate images [64], understand video content [32, 65, 82, 100, 123], and interpret data visualizations [108, 109, 115].

Prior studies have examined the strengths and shortcomings of these AI-driven applications in everyday activities [51, 52, 133]. For instance, González et al. [51, 52] conducted two diary studies with blind participants: the first documented common use cases of an AI scene description system but noted that the responses lacked depth and interpretive nuance; the second showed that richer LLM-generated descriptions directly shaped participants' trust, satisfaction, and willingness to adopt these tools in everyday contexts. Xie et al. [133] examined Be-My-AI in personal and social contexts and proposed *AI deferral learning*, where systems learn from the responses of remote sighted assistants to improve reasoning about socially sensitive judgments.

Beyond dedicated assistive tools, researchers have studied how blind individuals use general-purpose AI-tools such as ChatGPT. Adnin et al. [6] studied the use of AI tools in daily activities, focusing on how users navigated inaccuracies and hallucinations. Tang et al. [120] highlighted that blind users used these tools for content creation, PDF summarization, and visual interpretation, valuing the support but remaining cautious about their reliability. Trust remains a central concern: some blind users over-trust AI-generated descriptions [90], while others adopt more skeptical approaches as limitations become clearer [6, 11, 51, 63]. Expanding this line of work, Alharbi et al. [11] conducted an in-depth qualitative study with 26 participants, showing how users detect and interpret errors in AI-enabled assistance. Their findings revealed the non-visual strategies employed to verify information and underscored the need for accessible, contestable forms of explainable AI, offering concrete directions for Responsible AI grounded in disability studies.

Collectively, prior studies highlight both the benefits and limitations of AI tools in supporting blind individuals in activities such as navigation, social interaction, and productivity. However, the effectiveness of these tools in DIY tasks involving tangible products remains underexplored, in part because such tasks require multifaceted spatial reasoning, structural understanding, and cross-referencing diagrams or labels with physical parts, while support manuals primarily present visually-oriented instructions. Our work aims to address this gap.

### 2.2 Non-Visual Access in Everyday Tasks

Independence in everyday tasks is essential for blind individuals, but vision loss limits IADLs such as cooking and household management [20, 43, 78]. Cooking has been one of the most studied IADLs [24, 67, 74, 83, 84, 87], with prior work exploring non-visual recipe access through screen readers [14, 53, 77, 80, 111], braille displays [5, 33, 112], speech synthesizers [28, 57, 117], and tactile, image, or video formats [59, 65, 88, 94, 106, 129, 135, 138]. Even with these advances, studies report significant difficulties. A survey of 101 blind participants revealed that 65% of the blind users faced vision-related difficulties with cooking [67]. Reliance on visual cues hinders planning and execution, often pushing individuals toward packaged or restaurant meals with poorer nutrition [24, 74, 83]. Li et al. [87], through interviews with blind cooks and instructors, found that participants preferred concise step summaries and sensory cues such as auditory, olfactory, and tactile signals. Huh et al. [65] extended this work by proposing Vid2Coach, a smart-glass based solution that transforms how-to videos into accessible guides, monitoring participants' actions in real time.

Beyond cooking, prior work has examined other IADLs such as home management, laundry, and even personal care activities such as applying makeup [19, 43, 78, 85, 86]. For instance, Bhat et al. [19] investigated experiences of blind and low vision individuals with mobile and service robots. They identified accessibility challenges across the interaction lifecycle, from initiating and monitoring to controlling and stopping. Fine-grained control was particularly difficult because feedback often relied on visual outputs like blinkers or colored lights. These observations echo broader difficulties of operating everyday appliances when interfaces provide few or no non-visual cues [12, 121], including touchscreens without audio feedback, indistinguishable buttons, and sliders or knobs that lack clear state indication [13, 121]. Alternative approaches such as voice input [2, 17], gestural controls [36, 69, 125], and tactile augmentations or sensor-based monitoring [27, 72, 126] have been explored to help mitigate these barriers by providing non-visual channels for control and feedback. A common aspect of all these prior studies is that they are predominantly centered around the accessibility of visual system feedback (e.g. blinkers, colored lights, touchscreens) involved in everyday tasks. In comparison, product-manual-mediated DIY tasks involve challenges that extend beyond visual accessibility, e.g., the interpretability and operationalizability of manual instructions made accessible by AI and assistive tools.

---
[5]https://aira.io/

Furthermore, compared to previously examined tasks (e.g., cooking) where steps are flexible, familiar, and can be reordered, product-manual-mediated DIY tasks pose unique challenges. For instance, they are totally instruction-dependent, often rely on augmentative diagrams [7, 21, 35, 61, 101, 110], and require precise part identification, orientation, and alignment, where errors can halt progress or cause lasting damage. A recent prior work [92] studied how blind users assembled multi-part 3D-printed objects under different levels of assistance (i.e., unassisted, human-assisted), however the focus was limited to controlled and isolated user interactions with custom 3-D printed components; they did not examine the use of AI tools and product manuals by blind users in real-world assembly tasks involving actual products. The role of AI tools in product DIY tasks for blind users therefore remains underexplored, particularly the strategies they adopt and the barriers that constrain independent success; a gap this study addresses.

## 2.3 Assistive Technologies for Non-Visual Document Access

Instruction manuals are commonly available in printed form or as PDF files. A plethora of research has explored how to make them accessible to blind users, employing approaches such as OCR [31, 50, 93] for print and screen reader–compatible transformations for digital documents [38–40, 107, 127].

Early approaches to printed document access relied on sensory substitution to convert text into auditory or tactile signals. For instance, Optophone [31] mapped letters to musical chords, and the Optacon [50] translated characters into vibrotactile feedback. Recent advances in computer vision and OCR have enabled systems to scan printed documents, recognize text, and read them aloud through synthesized speech for blind users. SARA CE [49] and KNFB Reader [98] leverage these capabilities to provide non-visual access to printed materials, while wearable devices such as OrCam [104], Access Lens [68], Finger Reader [116], Magic Finger [136], Viet. et al[124] and Hand Sight [119] augment them with gesture or finger-based interaction. Applications such as Seeing-AI [93] and Be-My-AI [18] leverage VLMs to not only read printed content but also support interactive question answering.

Beyond printed formats, PDFs are a common medium of instruction manuals. Blind users access them with a range of assistive technologies, including braille displays [58, 134, 137] though screen readers [48, 103] remain the most widely used assistive technology[122, 127]. Many PDFs remain partially or entirely inaccessible [127] due to mismatches between the logical reading order and the sequence parsed by screen readers, or because they consist solely of scanned images [99, 122]. Although standards such as ISO 14289-1 [41] and PDF/UA (Universal Accessibility) [42] outline how to produce accessible PDFs, their complexity and the required in-depth understanding of accessibility considerations have hindered their widespread adoption [70]. Prior work has addressed both the creation of accessible PDFs [38–40, 107, 127] and the remediation of existing PDFs to make them usable for blind readers [75, 122, 128].

Prior work has also investigated how blind users access and interpret visual content such as images, charts, and infographics in documents. Adnin et al. [6] reported that blind users increasingly adopt AI tools such as ChatGPT for interpreting images and documents. Kim et al. [71] explored the potential of question-answering systems in supporting visualization interpretation and exploration for blind users. Complementary work on tactile charts and spatial layouts further demonstrates how non-visual modalities such as tactile graphics and audio-tactile maps support understanding of complex visual information and spatial structure [37]. Additionally, researchers have also developed several systems for making these visual content accessible for blind users [9, 54, 62, 115, 139].

Collectively, prior work has produced a range of general-purpose assistive tools that enable blind users to access printed and digital content more independently and efficiently. While these tools can render product manuals *technically* accessible, their *reliability* and *effectiveness* in supporting complex end-to-end task-oriented interactions, particularly for product-related DIY activities, remain largely unexamined. Specifically, it is unclear to what extent accessible instructions generated by assistive tools from product manuals are interpretable, actionable, and operationalizable by blind users in real-world contexts. Furthermore, little is known about how blind users currently leverage AI tools alongside product manuals for DIY tasks, and where existing systems fail to deliver dependable, non-visual, step-by-step guidance. Our work addresses these unresolved questions through a systematic investigation.

# 3 Uncovering Blind Users' Experiences With Product Manuals for DIY tasks

To investigate the challenges blind users encounter with instructions in product manuals for DIY tasks, we conducted an IRB-approved interview study.

## 3.1 Participants

We recruited participants from two sources: (i) a contact list maintained by the research team from prior IRB-approved accessibility studies, and (ii) the team's academic and professional networks. The contact list drew on mailing lists of regional and national blindness organizations and comprised participants who had opted in to be contacted for future studies. Additionally, snowball sampling [102] was employed, inviting enrolled participants to share study information with peers who met the inclusion criteria. Interested individuals were contacted via their preferred communication method, and eligibility was verified through a brief screening.

Participants were required to meet the following eligibility criteria: (i) be legally blind and fluent in English, (ii) be familiar with both screen readers and AI tools, and (iii) have prior experience using product manuals for DIY tasks. To mitigate potential confounds, individuals with mild visual impairments (e.g., low vision) who could employ residual sight or visual aids (e.g., hand-held magnifiers, screen magnifiers), and those under 18 years of age or without prior DIY manual experience were excluded.

| PID | Gender / Age | Vision Loss (Onset / LP) | Frequency of Manual Use | Screen Readers (Mobile/Desktop) | Tools used for Product-Manual-Mediated DIY tasks |
|---|---|---|---|---|---|
| P1 | M / 45 | 18 / Yes | Always | VoiceOver / JAWS (E) | Seeing-AI, Be-My-Eyes, Be-My-AI, AIRA, ChatGPT, Gemini (E) |
| P2 | F / 37 | 3 / Yes | Always | VoiceOver / JAWS (E) | Seeing-AI, Be-My-Eyes, Be-My-AI, ChatGPT, Gemini (E) |
| P3 | F / 52 | Birth / Yes | Always | TalkBack / JAWS (I) | Seeing-AI, Be-My-AI, Be-My-Eyes (B) |
| P4 | M / 64 | 6 / Yes | Always | VoiceOver / JAWS (I) | Seeing-AI, Be-My-Eyes, Be-My-AI, ChatGPT (I) |
| P5 | F / 66 | Birth / Yes | Sometimes | VoiceOver / JAWS (E) | Seeing-AI, ChatGPT, Be-My-AI, Be-My-Eyes, Lookout (I) |
| P6 | M / 61 | 18 / No | Always | VoiceOver / NVDA (I) | Seeing-AI, Be-My-Eyes, Be-My-AI, Lookout (B) |
| P7 | F / 51 | 19 / Yes | Always | VoiceOver / JAWS (I) | Seeing-AI, Be-My-AI, Be-My-Eyes, AIRA, KNFB Reader (B) |
| P8 | M / 59 | 28 / No | Sometimes | VoiceOver / JAWS (B) | Seeing-AI, Be-My-Eyes, Be-My-AI, ChatGPT, Gemini (I) |
| P9 | M / 60 | 17 / No | Always | VoiceOver / JAWS (I) | Seeing-AI, Be-My-Eyes, Lookout, Be-My-AI, ChatGPT, Gemini (I) |
| P10 | M / 66 | 5 / Yes | Always | TalkBack / NVDA (I) | Seeing-AI, Be-My-AI, Be-My-Eyes, KNFB Reader, Aira (I) |
| P11 | F / 58 | 28 / No | Sometimes | VoiceOver / NVDA (I) | Seeing-AI, AIRA, ChatGPT, Be-My-AI, Be-My-Eyes (I) |
| P12 | M / 43 | 5 / Yes | Always | VoiceOver / JAWS (I) | Seeing-AI, Be-My-Eyes, Be-My-AI, KNFB Reader, ChatGPT (I) |
| P13 | F / 68 | 19 / No | Sometimes | VoiceOver / JAWS (B) | Seeing-AI, ChatGPT, KNFB Reader, Be-My-Eyes, Be-My-AI (B) |
| P14 | F / 33 | 17 / Yes | Always | VoiceOver / JAWS (I) | Seeing-AI, Be-My-Eyes, Be-My-AI, ChatGPT, AIRA (I) |
| P15 | M / 73 | Birth / No | Sometimes | VoiceOver / JAWS (E) | Seeing-AI, Lookout, VDScan, Be-My-Eyes, Be-My-AI (B) |

Table 1: Participant demographics and technology use. Gender, age, and vision loss details (onset and light perception) are shown alongside frequency of manual use for DIY tasks (Always = manuals consulted for every task; Sometimes = manuals consulted selectively, with alternatives such as sighted assistance or tutorials often used instead), screen readers and tools employed, and participants' self-reported expertise levels: Beginner (B), Intermediate (I), and Expert (E).

Fifteen participants (8 male, 7 female) aged between 33 and 73 (M = 55.7, SD = 11.9) were recruited. All participants had prior experience with AI-assistive tools, primarily leveraging them for obtaining visual descriptions or reading printed text. Nine participants also reported using generative AI tools (e.g., ChatGPT, Gemini), for summarization or general information seeking. None reported additional disabilities limiting screen reader or AI tool use. Table 1 summarizes demographics, manual usage practices, and assistive technologies use.

### 3.2 Interview Design

The interview design was grounded in *a priori* tasks of interest, namely DIY activities involving tangible products. The questions were meticulously crafted to elicit concrete, first-person accounts, drawing on the principles of semi-structured interviews [4] and the recent critical incident approach [46]. The critical-incident approach enabled participants to reflect specific experiences, and provide richer, more detailed insights. They were encouraged to give detailed accounts of one product they had assembled, one they had learned to use, and one they had troubleshot in the last two months using product manuals. For each incident, interviewers documented the experiences including barriers, strategies or workarounds, and outcomes.

Prior to the sessions, the team meticulously evaluated the design for clarity and cognitive simplicity (e.g., concise language). After the first five interviews, the transcripts were analyzed using initial coding [30] to identify emerging patterns. These codes were then grouped into broader categories, which informed the subsequent refinement of the interview design and facilitated the development of more targeted follow-up questions in later sessions. This iterative process aligns with standard qualitative practices in HCI research involving blind participants [10, 23, 26]. The finalized protocol consisted of four main topics, structured with essential core questions and optional prompts to encourage further discussion. The core topics (along with seed questions) were:

(1) *Usage of Product Manuals (RQ1):* What kinds of DIY tasks have you used manuals for? What kinds of manual have you encountered? What modalities and layouts of manuals do you usually prefer? Why?
(2) *Accessing the instructions in Manuals (RQ1):* What kind of assistive technologies do you use to access manuals? What AI tools do you use to access manual instructions? How do you decide which AI tools to use when working on a DIY task? Do you have any preference regarding AI tools, depending upon the task?
(3) *Challenges in Following Instructions (RQ2):* What has your overall experience been like when trying to follow instructions from manuals and AI tools? When these issues arise, what do you usually do to move forward? What kinds of workarounds do you use? Did you often get stuck at certain points? If so, when? To what extent could these challenges be handled independently, and how often was that the case?
(4) *Suggestions for improvement (RQ3):* What changes or improvements would you like to see in the way the instructions are presented for DIY tasks?

### 3.3 Semi-structured interviews

The in-person interviews were conducted in English, each lasting between 45 and 60 minutes (M = 52). The interviewer summarized the consent form highlighting the study goals, confidentiality practices, and study duration. Participants were informed that they could skip any question or end the interview at any time. Written consent was obtained and, with explicit approval, the interviews were audio-recorded on Zoom. The interview began with demographic and background questions, such as age, gender, onset of vision loss and prior experience with screen readers and AI-based assistive tools. The remainder of the interview focused on four key topics (outlined in Section 3.2), each designed to explore a different dimension of participants' experiences with the DIY tasks. Although these topics guided the discussion, the follow-up questions were driven by the participants' accounts of the manuals they had encountered, the accessibility challenges they faced, and the strategies they used to overcome them. The participants received $80 as compensation for their participation.

### 3.4 Data Analysis

After all interviews were completed, each session was transcribed, yielding 734 single-spaced pages (maximum 50 lines/page). The transcripts were analyzed using hybrid reflexive thematic analysis [47, 97] combining deductive (top-down) and inductive (bottom-up) coding [25]. The deductive codes were developed from research questions and interview design, encompassed several low-level categories such as triggers for manual interactions and comprehension barriers. The first author reviewed the transcripts line-by-line, generating inductive codes by open-coding noteworthy responses. Deductive and inductive codes were merged into a unified codebook, reviewed by co-authors alongside transcripts, and refined collaboratively through weekly discussions. Whenever new codes were identified, the first author revisited all transcripts to ensure consistency. Final codes were grouped into higher-order categories to analyze relationships and derive broader insights. Coding and theme development were interpretive, shaped by the team's perspectives and expertise in accessibility research.

## 4 Findings

### 4.1 RQ1: Patterns of Manual Usage among Blind Users in DIY Tasks

*4.1.1 Blind Users' Reliance on Product Manuals for DIY Tasks.* Blind users designated product manuals as the *official* and *most trustworthy source of information* for DIY tasks. Ten participants said they consistently turned to manuals for assembly, setup, or learning new features, while the remaining five used them selectively depending on familiarity with the device or the availability of sighted help. Even when AI tools (such as ChatGPT), online tutorials, or sighted assistance were available, manuals remained *the primary resource*; other sources were treated as complementary rather than replacements.

Blind users did not undertake all assemblies indiscriminately; the complexity and perceptual demands of a product strongly influenced their choices. Most (n=12) participants preferred products with a limited number of components (about 10–15) that were easily distinguishable by touch. Larger items such as clothing racks and dressers were generally avoided. Several (n=10) participants favored products with minimal screws. Participants also preferred assemblies whose components could be fully laid out within reach. Learning and troubleshooting tasks, by contrast, did not impose the same physical constraints on what they were willing to attempt. Table 6 (Appendix A.1) lists the DIY tasks participants reported performing using product manuals.

The extent of reliance on manuals also depended on the type of DIY task (Figure. 2a). For *sequence-dependent tasks* such as assembly and setup, where mistakes could result in functional failure or safety hazards, manuals were described as *indispensable*. Ten participants said they were willing to spend extra time navigating a manual if it helped them "get it right the first time". For *learning to operate* new devices, manuals served as a supplementary reference when trial-and-error fell short. In contrast, for *troubleshooting*, manuals were rarely used. Participants described them as *high-effort, low-reward*, often offering generic advice that did not address the specific issue. Six participants even noted that troubleshooting sections were sometimes missing entirely. Several also mentioned that manuals were easy to misplace or discard after initial setup, further reducing their usefulness once problems emerged.

*4.1.2 Manual Modalities and Layout Formats.* The participants reported encountering manuals in three main forms: paper, digital (PDF or online), and audio. Paper manuals were the most

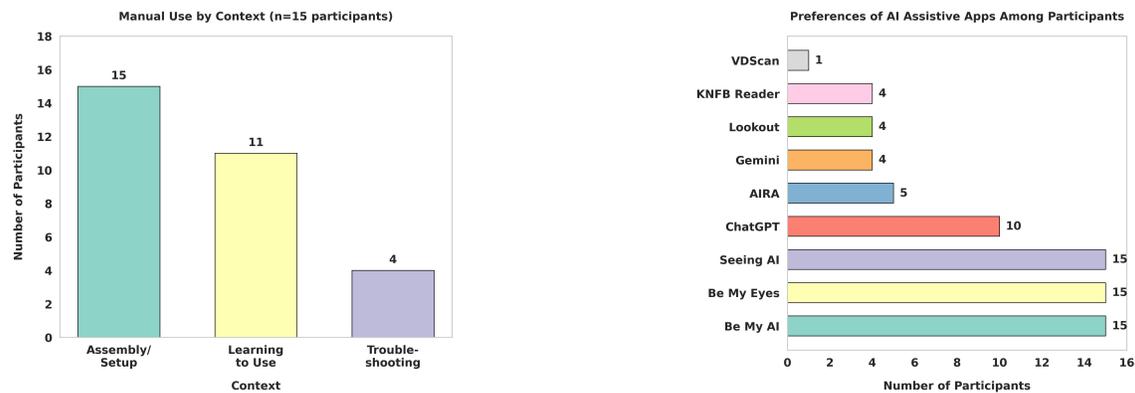

(a) Reported contexts of manual use among participants.

(b) Preferences of AI assistive apps among participants.

Figure 2: Reported contexts of manual use and AI assistive app preferences among the participants in our study.

common, digital versions were usually accessed only when paper copies were unavailable or misplaced. A few participants encountered audio manuals and described them as the most convenient format, but noted that they were rarely provided with everyday products. No participant in our sample encountered video manuals supplied by manufacturers.

Although *paper manuals* required capturing pages with AI assistive applications such as Be-My-AI and Seeing-AI, blind users still considered them the *most reliable format*. PDF and online manuals presented well-known accessibility challenges, such as mismatches between visual reading order and screen reader parsing, image-only files, and intrusive ads. Participants also faced new burdens: navigating these manuals often required constant switching between manual and assistive apps or AI tools on smartphones, a back-and-forth many described as frustrating and inefficient. Some (n=6) reported that uploading entire PDFs to AI tools like ChatGPT often produced irrelevant output, while page-by-page uploads improved accuracy but were laborious. Others (n=4) avoided PDFs or online manuals entirely.

Blind users strongly *preferred single-column layouts*, which minimized navigation complexity and supported independent use by allowing AI tools to interpret content correctly. Manuals with multiple columns or fold-out layouts were often set aside, with most participants (n=13) opting to wait for sighted assistance or search for online videos. P3 shared, "*I started following the steps, but after a while, it seemed to mix up two columns. I realized this because at least four intermediate steps were missing.*" Bilingual layouts posed similar challenges. P8 elaborated, "*...I tried capturing a part of English instructions, but it also picked up something in Spanish, and I couldn't make sense of what it was saying.*" In digital manuals, layout further complicated interaction by requiring frequent switching between files and assistive apps, fragmenting task flow.

### 4.2 RQ1: Approaches Blind Users Adopt in Accessing and Interpreting Product Manuals

Blind users preferred *smartphones for DIY tasks* as they contained the assistive tools required to scan pages, retake photos, and easily switch between apps. Three participants also experimented with smart glasses using Be-My-Eyes to read paper or PDF manuals, while the rest saw its potential for DIY use. For online manuals, screen readers were standard, and two participants converted digital manuals into braille, finding tactile formats easier to manage than frequent app-switching or digital layout inconsistencies.

No single AI tool was trusted to provide complete instructions. Blind users often combined AI-assistive-apps like Seeing-AI, Be-My-AI, and Lookout with general-purpose AI tools like ChatGPT, sometimes adding human-in-the-loop services such as Aira. The combinations depended on the previous experience. Figure. 2b shows participants' AI-assistive app preferences. Only three participants relied entirely on a single app (Be-My-AI), with prior positive experiences reinforcing confidence. P15 said, "*Sticking with a single tool is much easier than juggling three...It works well for me other tasks.....I believe it will work fine for this too..*" Twelve participants started with one app but switched when instructions were incomplete or confusing, moving opportunistically between tools guided by prior experiences. Only two participants engaged in deliberate cross-checking, capturing the same page with multiple apps and acting only when results aligned. While these approaches increased confidence, they did not always guarantee completion, and some participants reported abandoning tasks when conflicts persisted and repeated checks became overwhelming. Human support was a last resort, used when AI tools failed. Be-My-Eyes was preferred over alternatives like Aira. P10 explained, "*I'll try Seeing-AI and Be-My-AI first, and if it still doesn't make sense even after I ask more....then I switch to a volunteer. It's right there in the same app and that makes it much easier.*"

The adoption of new tools was also shaped by social influence and community knowledge. Blind users experimented with applications only after hearing about them from peers or online forums. P8 explained, "*Many AI tools continue to come out, it is hard to keep track of all... That's why we tend to stick with those that someone else has first experimented with. That's how I got my combo...*" Similarly, P11 recalled switching from Envision to Be-My-AI after learning about it in a community discussion.

## 4.3 RQ2: Challenges Blind Users Encounter with Manual Instructions in DIY Tasks

Despite using AI-assisted tools and general-purpose AI such as ChatGPT, participants found instructions often lacked sufficient non-visual guidance. They turned to YouTube or sighted helpers as a last resort, though these workarounds had limitations. Below, we outline persistent challenges of AI tools and supplementary strategies participants used to bridge these gaps.

*4.3.1 Inadequate Spatial and Orientational Guidance.* Before DIY tasks, blind users explored products by touch to identify each part, establish its orientation, and determine its function, a step they saw as essential for safe and accurate manipulation. However, they noted that this strategy often broke down when combined with instructions from AI-assistive tools or general-purpose AI tools (e.g., ChatGPT).

Regardless of the DIY task, participants (n=12) reported that instructions in manuals and AI tools extensively relied on *visual referents* such as labels (e.g., Part A), abbreviations (e.g., AL1), figure callouts (e.g., as shown in Fig.2), or colors (e.g., red button). Instructions often *lacked details on component orientation and spatial relationships*. Blind users were unsure how the device should appear assembled, which side was front, back, top, or bottom, or how parts related to each other. Attempts to infer orientation or adjacency from relative terms were inconsistent, shifting between egocentric and object-centric frames. Participants described being caught in a "*guessing game of perspectives,*" as P1 put it. Assemblies that seemed correct initially later revealed mis-alignments; P4 recounted assembling a Utility cart: "*At first it felt like the part fit, but when things started to feel off, I called a volunteer on Be-My-Eyes.. they told me it was wrong... I had to start over.*" Ad-hoc exploration became slow and error-prone; P2 described it as "*a sea of components,*" making even simple tasks, like setting a digital clock alarm, prone to mistakes. Attempts to clarify instructions via ChatGPT or the '*Ask More*' feature on Be-My-AI often yielded repeated or fabricated responses. Such cycles of dismantling, reworking, and verifying were time-consuming and demoralizing, turning even simple DIY tasks into prolonged, frustrating experiences.

As workarounds, participants often turned to online videos or sighted assistance. Videos relied heavily on visual continuity, with demonstrators pointing, gesturing, or showing on-screen text that screen readers could not interpret. Without descriptive narration, these cues collapsed for blind users, leaving only fragments of guidance. Sighted helpers, though well-intentioned, frequently used deictic references assuming a shared visual frame. The participants explained that with human-driven services such as Be-My-Eyes, the quality of support depended heavily on the individual. P6 recalled: "*... Be-My-Eyes volunteer kept saying, press that one...no, the other one....Without being able to see what she meant, I was just guessing.*" Several participants said they sometimes ended the call and reconnected to reach a more effective helper. Five participants reported that Aira's trained agents were slightly more helpful than volunteer-based services like Be-My-Eyes, but because it is subscription-based with only the first few minutes free, it was generally insufficient for completing entire DIY tasks. Overall, these workarounds, though occasionally helpful, often reinforced dependence on visual references rather than offering repeatable, tactile strategies.

*4.3.2 Ambiguity in Procedural Instructions.* Product manuals often use graphics such as arrows, icons, and exploded views to show actions (e.g., slide, twist), directions (clockwise or counter-clockwise), and step progressions. Sighted users can easily interpret these visual cues, however blind users rely entirely on AI tools. Participants reported that AI-generated instructions frequently missed these nuances, making tasks difficult to complete.

Participants highlighted that *vague phrases* like 'turn the knob' left them without the grounding needed to act confidently. Without explicit details on direction and extent, they were forced into trial-and-error, introducing uncertainty, wasted effort, and risk of device damage. P4 explained: "*The picture apparently showed arrows indicating the direction to turn the dial: clockwise for increasing, anticlockwise for decreasing the temperature. Be-My-AI said, 'turn the knob'. It never told me what the calibration of one turn - 5 degrees, 10 degrees or the direction? I had to guess everything, and I wasn't sure if I was even doing it right.*" Participants emphasized that instructions *often lacked confirmation cues* after each step, such as how the product should look, feel, or function after a step. P1 reflected: "*I never know if I've done it correctly.... They [sighted users] can just look and see if it's correct. I'm stuck.. wondering if I should move on or stop and wait for someone.*" Additionally, instructions *often failed to map part-labels to physical components via tactile features*, leaving blind users uncertain which part the instructions referenced. AI tools sometimes *hallucinated steps or components*, further complicating tasks. Several participants described wasting significant time '*chasing ghosts*' before realizing instructions were fictional. P2 recalled: "*It told me to press a button that wasn't even there. I kept running my hands over the surface, convinced I was missing something... It's frustrating...you start doubting yourself instead of the instructions.*"

Workarounds like YouTube tutorials often skipped steps or used jump cuts without narration. P7 said, *The person twisted the knob, I heard the click, but I had no idea what number he set it to.*" Participants often relied on videos created by blind users for clear step-by-step guidance, but such resources were rare. Some participants (n=9) noted that sighted helpers often assumed prior knowledge, giving instructions like "just do it like this" without explanation and merely confirming task completion rather than clarifying what made it complete.

Two participants (P4, P15) developed coping strategies to address these gaps: they rewrote the product manuals in Braille with the help of trusted sighted contacts and documented all non-visual cues. Other participants (n=7) avoided DIY tasks entirely, relying on the sighted assistance to complete them rather than face uncertainty and mental fatigue.

*4.3.3 Multi-action steps undermine task flow and error recovery.* The structure of the instructions in the product manuals created significant challenges for blind users. Instructions that combined multiple actions into a single step proved cognitively and practically onerous. Beyond the inherent difficulty of retaining several tasks in working memory while executing them, blind users faced three interconnected obstacles that together produced a cascading sequence of accessibility barriers.

Participants reported that multi-action instructions made it difficult to determine the next action in the sequence or pinpoint where they had gone wrong. Sighted users could visually scan instructions to spot a missed sub-action, but blind users often noticed errors only after completing several steps, wasting time and increasing frustration. This was worsened by linearized audio: screen readers delivered long continuous blocks, while AI-assistive tools sometimes repeated or reordered steps, obscuring where the mistake occurred. P12 shared her experience of setting up a printer: *"The manual said, 'insert the cartridge, peel off the protective tape and set blah blah.... I thought I had done it right, but I had missed pulling out the tape... The printer wouldn't start...I had to wait for my partner... I was frustrated... Turns out both the manual and Be-My-AI mentioned the tape, but it was buried in tons of information..."* Similarly, P6 described how uncertainty about the next action disrupted her ability to recover from errors: *"I knew I messed something up, but I couldn't tell where. It just says it all together. Sighted people can look and say, oh, I haven't fit that yet so let me do it and it will be fine. For me, I have no idea where I went wrong."*

Participants also reported a *cognitive burden* when they had to re-listen to entire multi-action steps, compounding confusion. Replays often repeated earlier sub-actions, making it difficult to track progress or isolate unfinished actions. P3 explained, *I go back to listen again, but it read the whole thing. This confuses me and I start doubting myself.. Have I done it or not..?"* This misalignment between execution and error detection led nine participants to abandon tasks and seek sighted assistance. Some (n=4) attempted to cope by externalizing instructions through braille notes or by asking sighted helpers to dictate steps, but even these strategies frequently broke down. P4 described, *'I try to write the steps out in braille, like one action at a time. But if the manual puts three things together, it's hard to separate them. I spend at least two hours just figuring this out."* Even sighted helpers struggled to ease the cognitive load. P7 recalled, *"They just read the whole sentence again... I had to ask them to repeat or read part of it. It's a little embarrassing."* Coping strategies offered little relief, highlighting how multi-action steps resist both individual and collaborative workarounds.

*4.3.4 Stitching Fragmented Information Together.* Although blind users relied on AI tools and YouTube tutorials, none of these sources provided reliable, end-to-end guidance for DIY tasks. Instead, the support was fragmented: each source offered only partial, and sometimes conflicting instructions. Participants explained that they therefore had to stitch together information from manuals, AI tools, and video tutorials–a cognitively taxing process that required them to track task state, remember conflicting instructions, and decide which source to trust. All participants emphasized that this process routinely stretched DIY tasks to *three or four times* the duration of their sighted peers. P6 illustrated this strain: *"I started with Be-My-AI, but it gave a bunch of instructions I couldn't follow... I paused halfway and asked ChatGPT what it meant, and it gave me something else... When nothing worked, I turned to YouTube and went through 10-15 videos, but they were all vague. By the time I finally found one that was detailed enough, I had completely forgotten what I was looking for or which step I was in..."* Even after piecing together fragments across different modalities, participants estimated that they succeeded without human intervention only about 10% of the time. They attributed this low success rate to the persistent absence of critical procedural, spatial, and feedback information across all forms of support.

As a result, blind users turned to sighted assistance. Although this support was usually reliable, it often came at the cost of independence because the task could no longer be completed on their own terms. The participants also noted that human help was not always available when needed, adding delay and dependence on others' schedules. Together, these accounts reveal a persistent tradeoff: participants could either preserve independence by performing exhausting orchestration work across fragmented sources, or sacrifice independence for the reliability of sighted support.

## 4.4 RQ3: Needs and Preferences

Participants highlighted that inaccessible instructions most hindered independence with household products (e.g., organizers, racks) and consumer electronics (e.g., thermostats, timers). These devices often have visually similar controls paired with manuals relying on diagrams and icons, forcing blind users into trial-and-error cycles. Here, we present the needs and preferences of blind users to improve instructions for DIY tasks.

*4.4.1 **Atomic steps with actions and confirmation**.* Participants emphasized that for blind users, 'step-by-step' instructions must present a single, complete action per step, rather than bundling multiple actions together. Each step should include four key elements: the action, its direction, a cue confirming successful completion, and for assembly tasks, the extent or magnitude of applied force. P13 illustrated: *"Rotate clockwise with light finger force until you feel one click; if there's no click, back off and realign."* Participants stressed that no detail is too small. Even minor cues help anchor orientation and sequencing, while measurable indicators such as clicks or degrees prevent misalignment, over-tightening, or incorrect selections.

*4.4.2 **Before–During–After Framing with Orientation Cues**.* Participants highlighted that instructions should specify the device's state before an action, the sensory change to expect during it, and the outcome signaling completion. This before-during-after framing was essential for maintaining a coherent mental model. Without it, users were unsure if they acted correctly or how their actions contributed to the entire task. P6 illustrated: *"Tell me, 'this is how it feels now.' Then say, 'you are supposed to do……..' After this, tell me, 'feel around, and this is how it should feel now.' That way I know I did it right."* Participants noted that this temporal scaffolding turned disjointed commands into a reliable narrative of progress, especially in assembly tasks.

All participants suggested including a one-line 'prelude step' before each step to establish a reference frame and resolve the deixis for the rest of the procedure. P8 illustrated: *"Tell me how to hold it first..logo facing me, hinge away. Then 'left' and 'right' finally mean something. Because if I get this wrong, everything after is wrong."* Participants also emphasized that any mid-procedure reorientation, such as rotating the device, should be explicitly mentioned as it reset the reference frame.

Participants noted that spatial relationships between controls and components must be conveyed through non-visual features such as tactile cues like beveled or square edges, hinge placement, surface texture, and relative positions such as below the big red button' or fourth button from the left'. When tactile cues were

| PID | Gender | Age | Vision Condition | Vision Loss (Onset / LP) | AI Tools used in DIY Tasks | Expertise |
| --- | --- | --- | --- | --- | --- | --- |
| P1 | Female | 37 | Retinitis Pigmentosa | 3 / Yes | Seeing-AI, Be-My-AI, ChatGPT, TalkBack | Expert |
| P2 | Male | 42 | Glaucoma | 18 / No | Seeing-AI, Be-My-AI, VoiceOver | Intermediate |
| P3 | Male | 61 | Retinopathy of Prematurity (ROP) | 5 / No | Be-My-AI, VoiceOver | Beginner |
| P4 | Female | 27 | Retinitis Pigmentosa | 19 / Yes | Seeing-AI, Be-My-AI, ChatGPT, VoiceOver | Expert |
| P5 | Male | 50 | Diabetic Retinopathy | 28 / Yes | Seeing-AI, Be-My-AI, TalkBack | Intermediate |
| P6 | Male | 31 | Retinitis Pigmentosa | 10 / Yes | Seeing-AI, Be-My-AI, VoiceOver | Intermediate |
| P7 | Female | 68 | Retinitis Pigmentosa | 3 / Yes | Be-My-AI, VoiceOver | Beginner |

Table 2: Participant demographics: gender, age, vision condition , vision loss onset and light perception (LP), tools used during DIY tasks, and self-rated expertise with assistive tools

insufficient, positional references relative to edges or expressed as clock-face positions were requested. Short exploration scripts anchoring flat interfaces or ports using landmarks, paths, and counts were also valued. P9 explained: "*When I'm setting a washing machine timer without tactile buttons, I need the instructions to say, "Turn the dial to six o'clock position for bedding." Then I know where it is and can set it independently*"

*4.4.3* **Error Recovery Paths**. Participants were frustrated that manuals assumed success at every step and provided no guidance when confirmation cues were missing or ambiguous. They emphasized that errors are inevitable in DIY tasks, especially when tactile or auditory signals are subtle, and wanted recovery instructions that were concrete and actionable rather than vague phrases like *if it doesn't work, try again*" or "*make sure it is aligned correctly*". P5 illustrated helpful guidance: "*If the part doesn't fit, don't just say 'try again.' Say, 'pull it out.. rotate it... press until you hear two clicks.' That tells me exactly what to do, and I won't have to guess.*" Without such contingencies, participants reported getting 'stuck in place,' often abandoning the task or seeking sighted assistance.

## 4.5 Summary

Blind users consider manuals indispensable for assembly/setup, but often abandon them for troubleshooting. Paper manuals, though requiring scanning, were viewed as more reliable than digital versions, with single-column layouts strongly preferred. Blind users drew on an assortment of resources, including AI tools, tutorials, and human support, yet these sources frequently provided fragmented and inconsistent guidance. Key barriers included reliance on visual referents, limited spatial cues, bundled steps, missing error-recovery, and inaccessible digital formats. To improve accessibility, blind users stressed the need for atomic step-by-step instructions with clear orientation, tactile descriptions, and explicit recovery paths and confirmations.

## 5 Usability Study

The semi-structured interview study (Section 3) highlighted the limitations of AI tools from the perspective of the blind users. To examine how these limitations manifest in practice, we conducted an IRB-approved usability study observing blind users performing DIY tasks with AI support in real time. This provided behavioral and procedural insights that interviews alone could not capture, surfacing nuanced challenges and failure patterns that hinder independent task completion and reveal opportunities for improvement.

### 5.1 Participants

Participants were recruited following the procedure outlined in Section 3. The inclusion and exclusion criteria remained the same, with minor modifications. To reduce variability, we only included blind users who regularly used product manuals in conjunction with AI-assistive tools for DIY tasks. Individuals who rarely used manuals or did not use them in conjunction with AI-assistive tools were excluded. Seven participants were recruited, with ages ranging from 27 to 68 years (M = 45.14, SD = 15.29). This group was different from the earlier semi-structured study, strengthening external validity. All participants were comfortable with Be-My-AI and Seeing-AI, and four had experience with ChatGPT and Gemini. None reported additional disabilities affecting screen reader or AI tool use. Demographic details are summarized in Table 2.

### 5.2 Design

The usability study combined concurrent think-aloud protocols [44] with in-situ interviews to identify challenges during real-time DIY task execution using AI tools. Participants were asked to complete tasks at their natural pace, and if they reached the allotted time limit, experimenters notified them and proceeded to the next task. They were instructed to only use the manuals and AI tools they normally relied on for DIY tasks. External aids such as video tutorials or volunteer assistance from Be-My-Eyes or Aira, were not permitted.

| Task ID | Task Category | Product | Manual | Task | Time taken by sighted user (min) |
|---|---|---|---|---|---|
| A1 | Assembly | Desk Organizer | Paper manual (1 page, 1 column) | Assemble the desk organizer (only Step 1 and 2) | 9 |
| A2 | | Ring Light | Web HTML manual | Assemble the ring light | 3 |
| T1 | Learning to operate & Troubleshooting | Timer | PDF (2 column, bilingual) | **Learning:** Set timer to ON between 7 AM and 7 PM. **Troubleshooting (Failure State):** Timer did not turn on, figure out why and fix it. | 6 |
| T2 | | Alarm Clock | Paper manual (fold-out, 1 page, 6 columns, trilingual) | **Learning:** Test if the clock is working and set the alarm for X AM/PM. **Troubleshooting (Failure State):** Alarm did not ring, figure out why and fix it. | 4 |

Table 3: Tasks used in the usability study: task categories, products, their OEM manuals (used as-is), assigned tasks, and baseline times recorded from sighted users.

Participants were asked to briefly verbalize their thoughts, actions, and to request assistance only when no viable progress could be made with the aforementioned resources. During the task, experimenters provided only the neutral prompt "keep talking" and avoided asking "why" or "how" questions. The experimenter intervened and provided guidance only when participants explicitly requested help, keeping in mind the needs and preferences uncovered in Section 4. Guidance primarily consisted of confirming whether a step had been completed correctly or clarifying how to carry out the current step. Each intervention was logged. Most of the participants verbalized continuously while performing the tasks and rarely paused. Within-participant comparisons showed similar pacing between verbalized and silent segments, indicating that verbalization had negligible impact on task completion.

Time limits were set at four times the mean completion time of six sighted users [6], based on the blind participants' accounts that they typically require three to four times longer to complete the tasks independently (Section 4). We used Williams Latin square [131] method to counterbalance task order and mitigate learning effects across both task categories and individual tasks. The workspace was equipped with a table, chair and non-slip mat. The study environment was kept consistent (lighting, noise, and tools). Between tasks, assembly products were dismantled and re-bagged, while the timer and alarm clock were reset to baseline and reseeded with faults for troubleshooting.

Participants completed four representative DIY tasks: two (A1, A2) focused on assembly and two (T1, T2) on learning to operate and troubleshoot using manuals that varied in modality, layout, and instructional style (Table 3). Guided by the findings from the interview study (Section 4), we selected products such that, across tasks, they aligned with blind users' preferences for product type, size, and operational complexity, and included Original Equipment Manufacturer (OEM) manuals whose formats corresponded to the ones they typically encounter in daily use. Table 3 summarizes the tasks, products, manual formats, and sighted-user baseline times. We employed a within-subjects design [56], with each participant completing all four tasks.

Tasks A1 and A2 focused on assembly. Task A1 (Desk Organizer[7]) involved 15 components, including 3 that felt similar to touch, differing only in minor features (e.g., small protrusions, pipe-like extensions). Instructions relied solely on diagrams for orientation and alignment, combining multiple actions into single steps; the first had three sub-actions and the second one, totaling four actions. Task A2 (Ring Light[8]) involved assembling five components that were easily distinguishable by touch, through four single-action steps, each specifying both what to do and how.

Tasks T1 and T2 focused on operation and troubleshooting. Task T1 (Timer[9]) involved a device with three controls: a dial, pins, and a switch. The dial markings and 'Time Now' arrow were purely visual, with no tactile features. After participants set the timer, the experimenter pressed random pins and switched to manual mode, creating a failure state. The manual included six single-action steps for setting the timer, but no troubleshooting guidance. Task

---

[6]To obtain a sighted baseline, we piloted all tasks with six sighted members of the research team using only the original product manuals, and the sighted completion times we report are the mean across these six users (See last column in Table 3).

[7]https://www.amazon.com/DeskEaze-Heavy-Wooden-Organizer-Drawer/dp/B0DT8MGN19?ref_=ast_sto_dp
[8]https://manuals.plus/bower/wa-trl10-tabletop-ring-light-kit-manual
[9]https://www.intermatic.com/userfiles/documents/tn300%20-%20instructions%20-%20en_es_fr.pdf

| Task ID | Product | Steps (As given in the manual) | Time Alloted (in min) |
|---|---|---|---|
| A1 | Desk Organizer | Step 1: Insert panels D into A, connect A and B with 4 pcs on M. connect A and C with 4 pcs on M Step 2: Install dividers E on panel B. | 32 |
| A2 | Ring Light | Step 1: Attach phone holder with pivot mount to ring light. Step 2: Screw ball head into bottom of ring light. Step 3: Attach tripod to ball head. Step 4: Secure smartphone in adjustable clamp. | 12 |
| T1 | Lamp Timer | Step 1: Pull all pins UP. Step 2: Rotate dial clockwise until "TIME NOW" arrow matches current time (White = AM, Dark = PM). Step 3: Push pins DOWN at desired ON/OFF times. Step 4: Ensure manual override switch is set correctly. | 24 |
| T2 | Alarm Clock | Step 1: Hold the Snooze and Time buttons for 3 seconds to enter TEST mode. Step 2: Within 1 minute, press Snooze or Time buttons to cycle through functions. Step 3: Press any button (except Snooze/Time) to exit Test Mode immediately. Step 4: Slide ALARM MODE SWITCH to AL1, AL2, or AL1+AL2. Step 5: Hold ALARM SET and press FAST/SLOW buttons to set alarm time. Step 6: Select 12h/24h format and adjust other features (snooze, buzzer, vibrator) | 16 |

Table 4: DIY tasks, step summaries, and estimated time to complete each task, covering assembly, installation, and electronic device setup/testing.

T2 (Alarm Clock[10]) had sliders, buttons, and dials. The manual used diagrams with numbered parts referenced by abbreviations elsewhere, requiring cross-referencing. After setting Alarm 1, the experimenter switched to Alarm 2 and partially disconnected the power supply, creating a failure state. The manual provided six steps for setting and testing the alarm, but no troubleshooting instructions. Table 4 summarizes the tasks, instructions, and allotted time for completing the tasks.

### 5.3 Procedure

The in-person study was conducted in English and lasted 110–140 minutes (M = 125). Two researchers were present: the lead researcher facilitated tasks, and a second monitored timing. None of the tasks posed a physical risk. At the start of each session, the consent form was read aloud and participants were informed they could stop at any time. Written consent was obtained and sessions were recorded using an iPhone, capturing only hands, the mobile screen, and interactions. The lead researcher also took detailed field notes on behaviors, reactions, and contextual observations. All videos and notes were securely stored on encrypted, password-protected drives. Participants first answered demographic questions and were introduced to the concurrent think-aloud protocol. The tasks were read aloud and all interactions with products, AI tools, and human interventions required were recorded. Before each task, participants had up to five minutes to familiarize themselves; timing began once they were ready. Participants accessed manuals via their preferred AI-assistive tools (e.g., Seeing-AI, Be-My-AI) on a their smartphone with screen reader support (VoiceOver or TalkBack). Five-minute breaks were provided between tasks. Each participant received $120 in compensation.

### 5.4 Metrics and Data Analysis

During the study, the lead experimenter collected the following quantitative metrics: (i) task completion rate; (ii) task completion time; (iii) step accuracy for each step in each task (without human intervention); and (iv) human intervention counts in each task. NASA-TLX questionnaire [60] was administered along with an exit interview after each task to capture participants' perceived workload. Qualitative analysis of the think-aloud utterances of the participants and feedback from the exit interviews followed the same coding and analytical procedures outlined in Section 3.

### 5.5 Results

*5.5.1 **Common DIY Strategies of Blind Users (RQ1)**.* Across all DIY tasks, participants followed similar strategies, which comprised the following:

***Getting overview of the manual***: All participants began by obtaining an overview of the manual content using their preferred AI tools. For paper-manuals (A1, T2), three participants (P2, P5, P6) used Seeing-AI, two (P3, P7) relied on Be-My-AI, and two (P1, P4) used both tools, cross-verifying their results for a comprehensive overview. For the web-based manual (A2), all relied solely on their screen readers. For the PDF manual (T1), three participants (P2, P3, P6) initially attempted to use screen readers, but because the PDF's reading order was scrambled and included mixed Spanish and English instructions, they switched strategies. They asked the experimenter to open the manual on a laptop and photographed the screen using their preferred AI tool on their smartphones. Four participants (P1, P4, P5, P7), already aware of common accessibility issues with PDFs, bypassed screen readers entirely: P5 and P7 captured images of the PDF on the laptop screen using Be-My-AI, while P1 and P4 downloaded the PDF to their phones and uploaded it to ChatGPT for a concise overview.

---
[10]https://www.amazon.com/Sonic-Alert-SBB500SS-Extra-Loud-Flashing/dp/B000OOWZUK?th=1

*Acquiring fine-grained details of the manual*: After getting an initial overview, participants sought more fine-grained guidance for each step in the product manuals. For this, five participants (P2, P3, P5, P6, P7) used Be-My-AI, while the other two (P1, P4) alternated between Seeing-AI and Be-My-AI, occasionally consulting ChatGPT to reconcile conflicting descriptions. They asked AI tools for part–label mappings, tactile, spatial, and orientation cues. Participants also used these tools for confirmations by uploading images (e.g close-ups of components, intermediate assemblies, full product). This was often a trial-and-error process: photos had to be retaken multiple times because of hand occlusions, poor focus, or misaligned framing before the AI produced a usable description. Drawing on their understanding of AI capabilities, participants expected these tools to stitch together the manual instructions, product as captured in their photos, and their current actions, to generate step-by-step guidance interpretable non-visually, enabling them to complete the DIY task independently.

*Prompting Styles*: During their interactions with AI tools, we observed three distinct prompting styles that varied depending on the participants' level of experience. Beginners often sent long, narrative 'everything-in-one-go' prompts, blending AI responses, their own observations, and requests. In A1, for example, P7 wrote: "You said B is a rectangular panel, now I can feel some wooden panel and many rectangular meshes with little circles and rods; can you please tell me which one I should use?". They often used part names differing from both the manual and AI instructions or relied on vague referents like 'those ones'. When participants could not understand the AI responses, they tried different prompts. Over time, their prompts began to include words that expressed their frustration. When Be-My-AI repeatedly used unclear part labels, P3 wrote: "I literally told you I don't know what A or B is, and you keep telling me that. What the heck is A and B?" Intermediate users relied on step-specific queries and avoided long, verbose prompts. In T2, for example, P2 first asked, "Where is AL1?" When Be-My-AI responded vaguely ("on the alarm set button"), they followed up with, "Where is the alarm set button? Can you describe Step 3 again with details on how to locate the alarm mode switch and the AL1 button?" Expert users adopted the same step-specific style but went further, reusing exact component descriptions or labels and adding explicit constraints: "Don't use anything that requires vision like colors. Give precise steps." Both intermediate and expert users probed AI tools to resolve ambiguity and obtain actionable guidance. Table 7 (Appendix A.2) presents actual prompts and AI responses illustrating these patterns.

*5.5.2 Task completion rate and Task completion time.* Tasks were considered complete if participants performed all steps correctly without human intervention within the allotted time. The completion rates for the tasks were: Tasks A1, T1 and T2 had 0%, while A2 reached 14%, resulting in an overall completion rate of DIY tasks of 2.4%. A2 had the highest completion rate, because its instructions were clear, precise, and composed of single-action steps, with minimal reliance on diagrams, and the parts were easily distinguishable by touch. This non-visual interpretability allowed participants to progress more accurately and make fewer errors.

Task completion time (the difference between task start and end time) was computed manually from the video recordings beginning when participants opened the packaging and ending at success, an explicit decision to stop, or upon reaching the allotted time limit. Three researchers independently logged times to the nearest second; reported values are their mean. All participants exceeded the allotted time on every task except P2, who completed A2 in 10 minutes out of the 12-minute limit.

Figure. 3 shows task completion time breakdown. For each task, we reviewed video recordings from when participants opened the packaging until task completion, withdrawal, or the time limit was reached. Three researchers independently marked the start and end of meaningful actions to the nearest second (e.g. AI prompts, AI responses product exploration). Periods of slowed progress or requests for human interventions were annotated with contributing factors (e.g. part-label mismatches, hallucinations, etc). For each factor, we summed all intervals where it was annotated for each participant, divided by the participant's total on-task time, then averaged across participants. Discrepancies in annotations and interval durations were resolved through discussions, with final verification by the co-authors.

Notably, AI-related failures (discussed in detail in Section 5.5.5) consumed a substantial portion of task time. Prompting style had little effect: beginners wrote longer queries and took a few seconds longer per prompt than more experienced users, yet composing prompts and waiting for responses accounted for less than 2% of total time across the tasks. Layout- and modality-related issues consumed ($\approx$ 10–15% of task time). In A1, T1, and T2, participants adjusted the camera and retook photos because of misalignment, hand obstruction, or poor focus. Seeing-AI and Be-My-AI sometimes flagged images as 'cut off' without suggesting how to correct them, resulting in trial-and-error. Five participants showed overt frustration and two considered abandoning the task before choosing to proceed. In A2, accidental clicks on pop-up ads repeatedly redirected participants away from the manual, making them relocate their current step. In T1, the PDF's scrambled reading order mixed English and Spanish instructions, so five participants chose to photograph the manual on the laptop screen instead. Human intervention accounted for only $\approx$ 4% of task time, and switching between AI tools added 5–10% for P1 and P4.

*5.5.3 Task Accuracy and Human Intervention Counts.* Given the strict step order required in DIY tasks, we measured task accuracy as the percentage of steps completed correctly without human assistance, with any error rendering all subsequent steps incorrect. Overall, task accuracy with AI-generated instructions remained below 50% ( see Table 5). Among the tasks, A2 yielded the highest accuracy: five participants completed the first two steps correctly, and one completed the entire task. Although the instructions were clear, pop-up ads in the web manual disrupted task flow, causing several participants to accidentally skip or reorder steps and thereby lowering task accuracy.

We also found no systematic differences in task accuracy across the different AI tools (used by P1, P4). Prompting style similarly had no effect: whether participants used long, verbose prompts (typical of beginners) or concise, targeted prompts (typical of more experienced users), task accuracy stayed uniformly low.

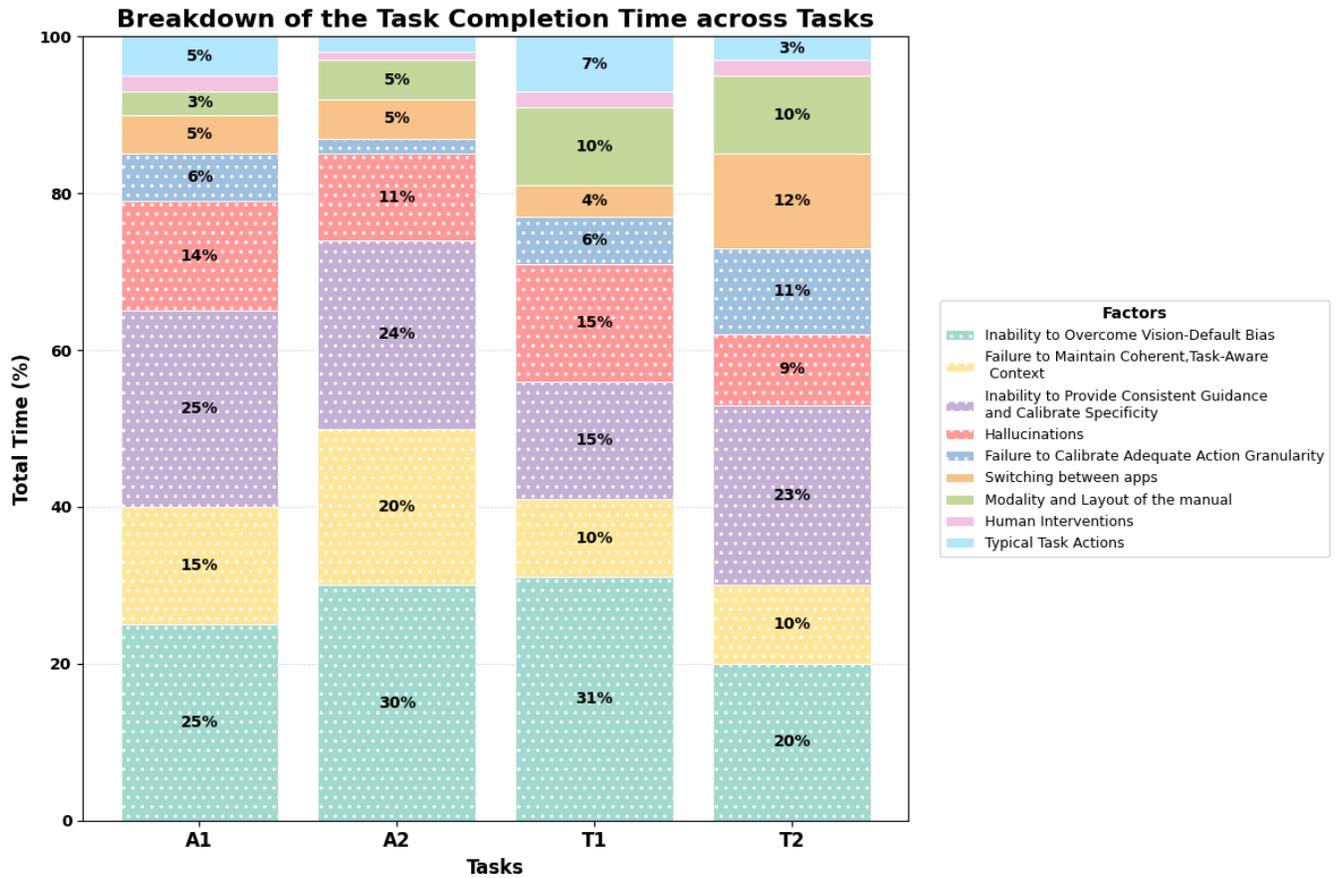

Figure 3: Breakdown of factors contributing to increased task completion time across tasks (A1, A2, T1, and T2). Bars show the proportion of the allotted task time attributed to each factor, with percentages indicating that factor's relative contribution within each task. Segments with white-dotted fill denote AI-related failure factors, whereas solid segments represent non-AI factors and typical task actions (e.g. waiting for AI response, exploring the components).

| Participant | A1 | A2 | T1 | | T2 | |
|---|---|---|---|---|---|---|
| | | | Learning | Troubleshooting | Learning | Troubleshooting |
| P1 | 0% (0/4) | 67% (2/3) | 0% (0/4) | 0% (0/5) | 0% (0/6) | 0% (0/7) |
| P2 | 75% (3/4) | 100% (3/3) | 0% (0/4) | 0% (0/5) | 0% (0/6) | 0% (0/7) |
| P3 | 0% (0/4) | 0% (0/3) | 25% (1/4) | 0% (0/5) | 0% (0/6) | 0% (0/7) |
| P4 | 0% (0/4) | 33% (1/3) | 0% (0/4) | 0% (0/5) | 0% (0/6) | 0% (0/7) |
| P5 | 0% (0/4) | 67% (2/3) | 25% (1/4) | 0% (0/5) | 0% (0/6) | 0% (0/7) |
| P6 | 0% (0/4) | 0% (0/3) | 0% (0/4) | 0% (0/5) | 0% (0/6) | 0% (0/7) |
| P7 | 0% (0/4) | 67% (2/3) | 25% (1/4) | 0% (0/5) | 0% (0/6) | 0% (0/7) |
| Average | 11% | 48% | 11% | 0% | 0% | 0% |

Table 5: Step accuracy without human intervention across tasks. Percentages are computed as steps completed correctly over total steps (shown in parentheses).

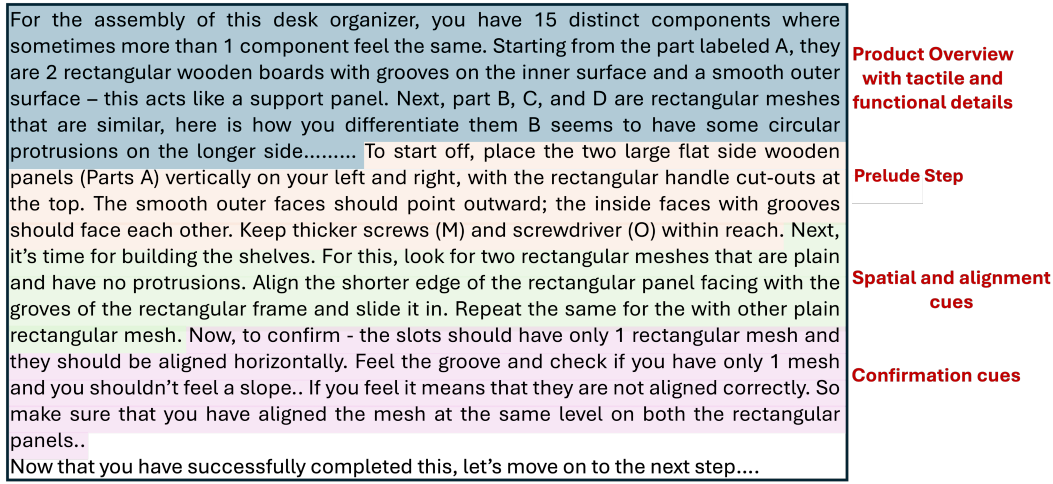

Figure 4: Snippet of the experimenter-participant conversation illustrating how DIY instructions were made understandable for blind users. The example (shown here for a desk organizer (task A1)) highlights four key layers: (top) a product overview with tactile and functional details to differentiate similar components, (next) an orientation preface (Prelude step) that establishes reference frames, (then) spatial and alignment cues guiding correct positioning, and (bottom) confirmation cues enabling users to verify successful completion. This layered approach ensures atomic, verifiable steps while reducing ambiguity and trial-and-error.

Only prior task-specific experience reliably improved accuracy. P2, a former carpenter, achieved 75% in A1 and 100% in A2, using his familiarity with parts and assemblies to compensate for inadequacies in the AI instructions. Participants with similar or greater AI-tool proficiency (e.g., P1, P4) but no comparable hands-on experience struggled with the same tasks. His expertise did not extend to T1 and T2 (learning and troubleshooting), suggesting that prior experience helps only when the DIY task draws on procedural skills the user has already mastered.

Human Intervention Counts indicate how often participants requested assistance from the experimenter. Each request prompted the experimenter to restructure instructions into a non-visual format informed by interview findings (Section 4). For overview questions, the experimenter described the product structure, mapping each part to its label and tactile features (e.g., "Part A refers to a rectangular wooden board……"), helping participants form a clear mental model of the product. For step-specific questions, instructions followed a consistent pattern: (i) a brief summary of the step and a prelude orientation, (ii) tactile cues for selecting correct components, (iii) spatial and orientation cues, and (iv) non-visual confirmation checks and error-recovery guidance (e.g, Each slot should fit only one smooth rectangular panel). Instructions were delivered incrementally and advanced only when participants said "Next." Figure. 4 illustrates this pattern. When presented like this, the number of re-clarifications reduced by nearly 80% compared to AI re-prompts, demonstrating that structured, incremental instructions effectively supported independent and accurate task completion.

5.5.4 **NASA-TLX**. We administered the NASA–TLX [60] to assess perceived workload across tasks. For blind users, DIY tasks require assembling, learning to use, or troubleshooting products while interpreting instructional content non-visually and interacting iteratively with AI tools and screen readers. These demands create cognitive and physical workload, as well as time pressure and frustration, which are not fully captured by prior objective metrics. Overall, NASA-TLX scores (Figure 5) indicated that A1 had the highest workload (M = 86.81, SD =4.03). T1 (M = 78.62, SD = 4.12) and T2 (M = 74.00, SD = 7.45) were moderate, though T2 showed greater variability. A2 was lowest (M = 40.57, SD = 6.40), clearly less demanding. A closer examination of the sub-scales revealed that mental demand and frustration were the primary contributors to these differences, while temporal demand remained relatively stable across tasks.

To better understand the NASA-TLX scores, participants evaluated the instructions on clarity, consistency, and guidance. Participants reported that inadequate non-visual cues in AI responses left participants without a clear mental model of the product and its parts, forcing them to mentally simulate each step in 3D, increasing the cognitive load. The specific AI-related factors contributing to this increased workload are discussed in detail in Section 5.5.5.

Participants reported that the non-visual instruction pattern in Figure. 4 'made sense' and could be followed without repeated clarification. By consistently identifying the parts, describing the actions, and stating the expected outcomes, the instructions established a clear rhythm that reduced ambiguity, supported orientation, and clarified task progress. This suggests that *instruction effectiveness depends not only on content but also on its organization and presentation*. Participants emphasized that this structured pattern enabled them to complete tasks more independently, and noted that if AI tools adopted a similar structure, tasks could be performed with less confusion and greater autonomy.

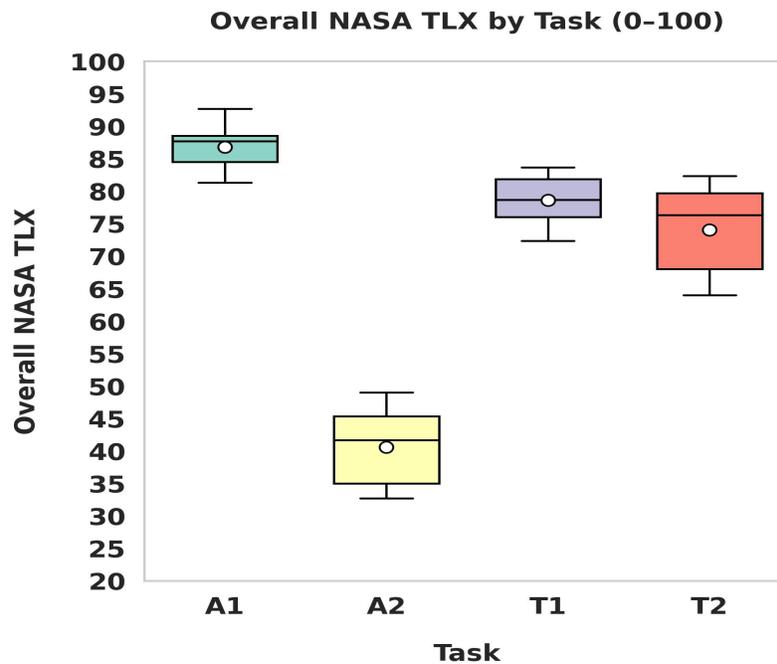

Figure 5: Boxplots of overall NASA TLX ratings for tasks A1, A2, T1, and T2. Higher scores indicate greater subjective workload. White circles mark the mean within each task.

*5.5.5 Factors Influencing Performance Metrics (RQ2).* The factors uncovered below highlight areas where current AI tools may still need improvement, as they at times made tasks harder, lowered task accuracy, and increased cognitive load, resulting in the need for human involvement.

*1. Inability to Overcome 'Vision-Default' Bias.* AI tools occasionally provided effective non-visual guidance (e.g., in A1, ChatGPT described Part A as a rectangular panel with an oval handle cut-out, and in T2, Be-My-AI identified the ALARM SET button as the first from the left), enabling participants to progress confidently. Such instances, however, were rare. Consistent with our interview findings, even when participants explicitly requested non-visual cues (e.g., I am blind; please tell me something I can feel and follow"), roughly 80% of responses still relied on colors, labels, or figure references. AI tools frequently omitted tactile, spatial, and orientation cues and rarely linked part labels (e.g., Part A") to tactile features that would help participants locate the corresponding physical components (Figures. 6 and 7). Consequently, participants repeatedly re-prompted AI tools for clarification, consuming 20–30% of task time (Figure 3). However, these follow-ups seldom improved guidance, leading to misidentified components, incorrect manipulation, task abandonment (e.g., P4 in A1), and increased reliance on human intervention. This pattern highlights a key limitation: *although AI tools can provide non-visual guidance, they have high tendency to default to vision-focused instructions, ignoring the actionable cues blind users require.* All participants noted: *We know the manuals are vision-dependent, hence we rely on AI... We expect it to map parts with labels and tactile information, like how part A feels, how to hold it... It gets frustrating when they just repeat the manual..."* This disparity between expectations and actual AI behavior directly resulted in elevated NASA–TLX ratings for mental demand, effort, temporal pressure, and frustration.

*2. Failure to Maintain Coherent, Task-Aware Context:* In our study, a counterintuitive pattern emerged: *uploading product photos frequently degraded guidance rather than enhancing it.* When blind participants uploaded product images to AI tools and posed targeted procedural questions, such as "Is this the manual over-ride switch?" (T1) or "Which of these panels corresponds to Step 1?" (A1), AI responses frequently exhibited *response misalignment*. For instance, AI-assistive tools often defaulted to an image-descriptor mode, emphasizing generic visual descriptions over answering the user's query (e.g., The image shows a person holding a timer). Instructions that followed were disconnected from the participant's current step and often tangential or irrelevant. This reveals a systemic limitation of current AI tools within the scope of complex, multi-step DIY tasks: *they fail to sustain a coherent, task-aware context and integrate uploaded product images, iterative user queries, and the actual manual content to provide actionable, non-visual stepwise guidance.*

*3. Inconsistent Guidance, Specificity, and Action Granularity:* AI tools occasionally provided consistent references (e.g., ChatGPT consistently identified Part D as 'plain rectangular flat panel' and Part B as 'rectangular panels with circular protrusions' in Task A1), but such cases were rare. More often, they referred to the same component by different names across turns (e.g., in A1, 'a rectangular mesh' was later called a 'shelf' and then a 'black panel'; sliders were called 'sliders,' 'buttons,' or 'tabs' in T2), prompting repeated clarification from participants. AI tools rarely decomposed multi-action instructions into sequential single-action steps. Essential orientation guidance (i.e., explaining how to hold, position, or align the parts) before each step was seldom provided, and explicit prompts like 'Tell me how to hold it' often elicited vague guidance such as 'attach it on the ring' or relational references such as 'a switch next to the snooze button.' Instructions were typically verbose multi-action steps rather than single-action sub-steps, causing participants to replay them 8–10 times per step. Participants were caught in a back-and-forth cycle, alternating between performing the task and consulting the instructions. Step accuracy also suffered, as they often selected the wrong component, oriented it erroneously, or missed critical steps. These issues also increased cognitive load and in some cases led to task abandonment (P4 abandoned task A1). This reveals a critical, task-specific limitation: *although AI tools could convey the required information, their guidance was often inconsistent, failed to maintain coherent terminology, misjudged the appropriate level of detail, and rarely decomposed multi-action instructions into clear, single-action steps that blind users could reliably follow for DIY tasks.*

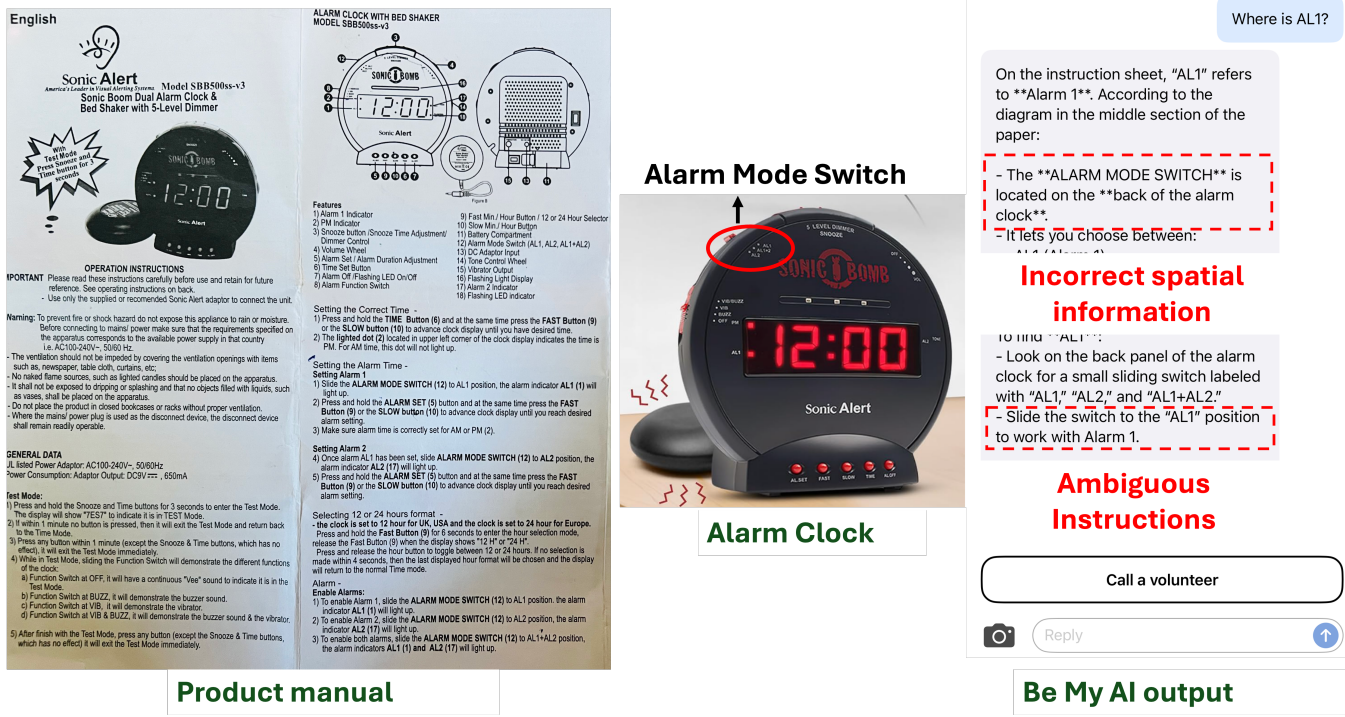

Figure 6: Conflicting gaps in AI support for Task T2: (Left) The product manual is a fold-out sheet with six columns in three languages; shown here is a photo of the two English columns. (Middle) The actual device confirms the switch is on the top, illustrating how AI errors in spatial descriptions can mislead blind users. (Right) Be-My-AI misdirected the user to look at the back panel, providing spatially inaccurate instructions.

**4. Hallucinations:** In 64% of all AI responses across participants, the AI tools produced confident yet hallucinated responses, instead of acknowledging uncertainty or suggesting human assistance. Specifically, we observed four main types of hallucinations [113]: (i) *Semantic hallucinations* mis-characterized how a component or interaction worked, leading participants to misunderstand intended actions (e.g., describing a non-existent lock mechanism between parts B and C in A1 or instructing participants to listen for a click that never occurred in T2); (ii) *Content hallucinations* introduced non-existent parts or controls, prompting participants to search for items that did not exist (e.g., two rectangular curved mesh panels in A1 or an AM/PM slider on T2); (iii) *Factual hallucinations* mis-stated concrete procedural details such as step order or tactile cues, often blocking progress until multiple steps were corrected (e.g., setting ON/OFF times before the current time in T1 or asserting the 'TIME NOW' arrow was raised when it was not); (iv) *Contextual hallucinations* shifted instructions away from the current step or contradicted prior constraints, producing tangential guidance (e.g., in task A1, when a participant asked how to attach panels A and D, the AI instead provided instructions for securing panel F); participants usually detected and recovered from these by re-prompting. Semantic, content, and factual hallucinations increased the need for human intervention, while semantic and content hallucinations also led to wasted effort and repeated, futile exploration.

These findings align with our interview study, emphasizing the challenges with AI-generated instructions for blind users, which can sometimes lead them to pause or abandon product-manual-mediated DIY tasks.

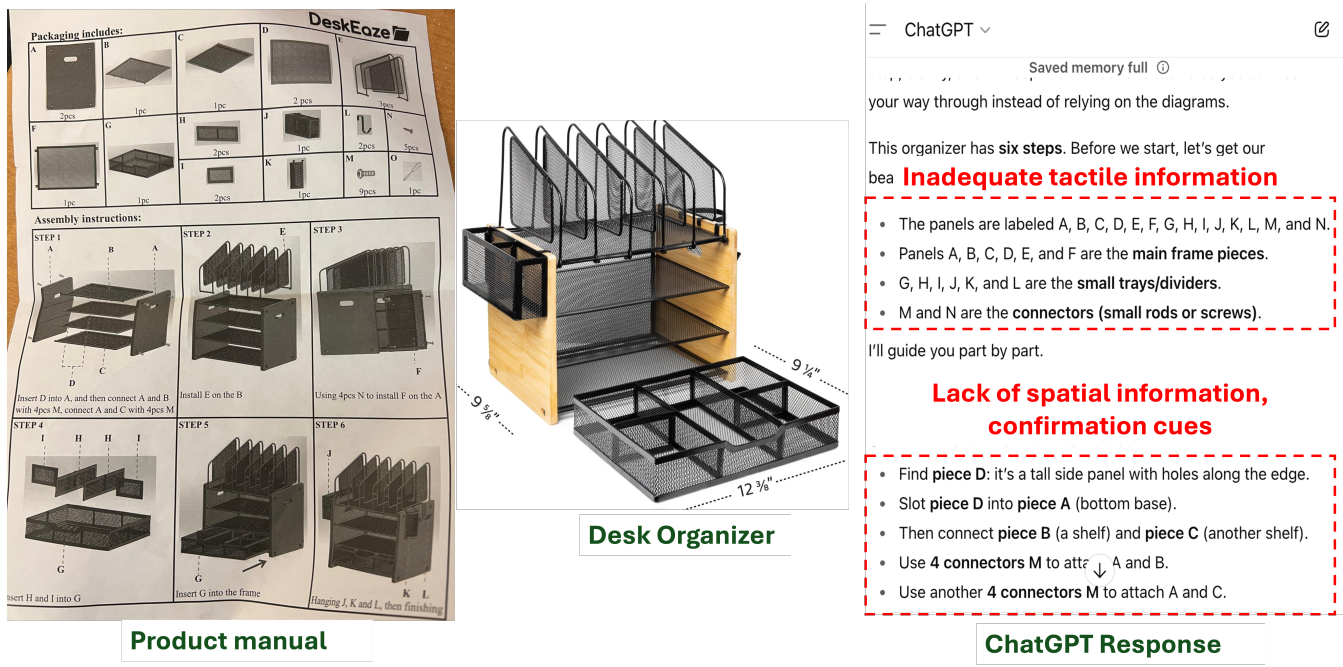

**Figure 7: Conflicting gaps in AI support for DIY tasks:** (Left) The product manual provides clear diagrams and step-by-step visuals for assembling the desk organizer. (Middle) The product. (Right) ChatGPT produced text-only instructions that lacked tactile clarity and spatial confirmation cues, leaving the user uncertain about panel orientation and connections.

## 6 Discussion

Our study uncovered persistent gaps in support provided by current AI tools to blind users for completing product-manual–mediated DIY tasks independently. Below, we situate these findings within prior work, outline design implications for future systems, and highlight our study limitations and directions for further research.

### 6.1 Advancing Understanding Beyond Findings of Prior Studies

*6.1.1 Identifying Challenges Unique to Product-manual mediated DIY Tasks.* We significantly advanced understanding of blind users' DIY practices with consumer products, spanning information seeking through physical task execution. Prior work had mainly examined blind people's use of AI tools in everyday activities such as navigation, cooking, shopping, and general information access [6, 32, 51, 52, 133], highlighting broad issues such as hallucinations, inadequate descriptions, difficulty interpreting contextual cues beyond visual appearance (e.g., cultural, situational, personal context), and trustworthiness. While these findings had surfaced important and conceptually similar challenges, DIY tasks impose distinct requirements. Unlike domains such as cooking or navigation, where steps are flexible and errors recoverable, product-manual-mediated DIY tasks required strict step order, precise part identification, correct orientation, and accurate alignment; a single missed step or misaligned part could halt progress or render the product unusable. Our study focused exclusively on product-manual–mediated DIY tasks: how blind users operationalized AI-rewritten instructions to assemble, operate, and troubleshoot consumer products essential for home management and independent living. In this context, we found that: (i) AI-rewritten instructions, though accessible, often lacked comprehensibility for precise execution because they omitted essential non-visual cues such as tactile references, confirmation checks, spatial relationships, and orientation guidance; (ii) even when AI tools provided useful non-visual guidance, it was inconsistent, as instructions frequently reverted to visually biased, multi-action steps that forced repeated re-prompting; (iii) even expert blind users' sophisticated prompts could not reliably sustain non-visual guidance over the course of a task; and (iv) blind users often had to stitch together fragmented information from product manuals, AI tools, and YouTube tutorials, a cognitively taxing process that frequently left tasks incomplete despite extensive effort. Our work also extended prior accounts of hallucinations [6, 51, 52, 133] by categorizing the hallucinations that arose in DIY contexts and identifying which ones impeded task completion. Context hallucinations rarely affected the task and were recoverable. Factual, semantic, and content hallucinations frequently derailed progress. Additionally, our findings on the limitations of video tutorials and sighted assistance echoed prior work [66, 132]: videos often relied on visuals with minimal narration, and sighted help was inconsistent in availability and quality. Taken together, these breakdowns revealed a deeper gap in contemporary AI support for blind users' independent completion of product-manual-mediated DIY tasks.

*6.1.2 Advancing Understanding of AI-Supported, Manual-Mediated DIY Workflows*. Our findings significantly extend understanding of the challenges blind users face when assembling, operating or troubleshooting consumer products with AI tools and product manuals. Prior work has predominantly focused on (i) assembling custom 3D-printed objects with and without human assistance, showing that success hinges on reliable part identification and correct orientation [92], and (ii) operating consumer devices, documenting interface-level issues such as visual-only feedback (e.g, blinkers or colored lights), indistinguishable flat buttons, and companion apps that work poorly with screen readers [12, 19, 121]. However, these studies did not consider how blind users interacted with product manuals or AI tools for these tasks. Consequently, it remained an open question whether their findings generalize to manual-mediated DIY workflows in which blind users rely on AI tools to convert vision-centric instructions into non-visual, step-wise guidance. Drawn from real consumer products and corresponding OEM manuals, our study corroborates that part identification, orientation demands, and interface barriers remain central obstacles. We further show that current AI tools are largely unaware of non-visual requirements: they echo manuals' visual referents instead of restructuring them into non-visual, operationalizable steps. Even if they do, they often lose context, provide inconsistent or hallucinated guidance, and bundle multiple actions into single steps—patterns that hinder independent DIY completion. Finally, we uncovered a characteristic two-stage workflow: blind users first use AI tools for a high-level overview of the manual and then issue step-specific prompts for tactile, spatial, and orientation details.

*6.1.3 Revisiting AI's Role in DIY Product-Manual Accessibility for Blind Users.* Prior work has documented document accessibility issues such as screen readers reading content in an order that does not match the visual layout and missing or inadequate alternative text for figures and tables [8, 15, 45, 91, 114, 118, 128, 140]. Complementary work has shown that AI-powered description and question–answering tools can help blind users interpret charts, infographics, images, and tables, and can summarize PDF documents [6, 9, 54, 62, 115, 139]. While these systems are valuable for exposing and explaining visual and document content, it is not known how well they support blind users in product-manual–mediated DIY tasks, which require strict step order, reliable mapping between textual labels and physical parts, and non-visual cues that support safe, accurate manipulation. Our findings show that, in this setting, AI tools can summarize and describe manuals, but they rarely generate instructions that can be followed in practice. AI-rewritten manuals frequently preserve vision-centric assumptions, omit essential non-visual details such as tactile, spatial, and orientation cues, and fail to organize information into clear, stepwise procedures. As a result, manuals and their AI-generated rewrites may be accessible at the content level but not operational for DIY use, since current AI tools do not reliably transform vision-centric manuals into non-visual, stepwise instructions that blind users can interpret and carry out independently.

## 6.2 Redefining Support for Blind Users' Independent DIY Task Completion

Our findings provide insights for tailoring support for blind users in assembling, operating, and troubleshooting products, and they point to two complementary directions. First, product instructions can be redesigned to better enable accurate, efficient, and independent DIY task completion through a clean, step-wise structure and reliable non-visual information. For instance, a single-column format with one action per step, paired with raised edge markers and tactile checklists, helps users stay oriented, reduces memory load, and supports error recovery. QR codes and object-linked cues such as NFC tags can further connect guidance to the components in hand, enabling part-specific support for orientation, assembly, and completion checks without requiring users to translate diagrams into objects. Table 8 (in Appendix A.3) presents these design recommendations in detail.

Second, AI tools can advance the same goals of clarity, grounding, and reduced cognitive load by generating guidance that is both accessible and operationalizable by blind users during these hands-on product-manual-mediated DIY tasks. This includes producing layered instructions that progress from overall component layout, to sub-component connections, and then to fine-grained tactile cues. Orientation guidance can be made more consistent by grounding it in user-relative language and identifiable features, drawing on spatially informed methods such as SpatialRGPT [34]. Prompt chaining [76], retrieval-augmented generation [81], and Few-shot prompting [29] could be combined with Chain-Of-Thought [130] to encourage step-by-step reasoning within a structured template (Figure. 4), providing clearer troubleshooting guidance. These recommendations are detailed in Table 9 (in Appendix A.3).

## 6.3 Limitations and Future Work

Our study had several limitations. First, our samples were small: the usability study included seven participants, only two to three of whom were beginners, and the interview study was similarly limited in size. As a result, our findings may under-represent the experiences of novice or less confident technology users. Second, all interviews were conducted in English, so we do not capture multilingual contexts or experiences of users who rely on non-English AI interactions. Third, most usability-study participants used Be-My-AI, which at the time of the study (June-July 2025) was powered by GPT-4; future models and interface changes may alter the types of instruction failures we observed. Finally, the lab-based usability study is subject to observer bias and the dual-task demand of performing tasks while thinking aloud. However, participants' relatively stable pacing suggests that the effects of the dual-task load were likely minimal.

Future studies could expand participant diversity, especially by including novices and non-English speakers, to test generalizability. Research should also evaluate a wider range of DIY manuals beyond those studied here to determine whether accessibility barriers differ across product categories. Longitudinal work could track how accessibility challenges evolve as AI models change. Another direction is to prototype and evaluate systems that implement layered templates, orientation reasoning, and action segmentation with blind and visually impaired users, providing empirical validation of these design interventions. Finally, exploring new delivery modalities, such as smart glasses could enable hands-free access to step-by-step guidance during physical interaction with devices.

## 7 Conclusion

Our studies investigated how blind users engage with product manuals and AI tools such as ChatGPT and Be-My-AI to assemble, operate, and troubleshoot tangible products. Through interviews and usability studies, we found that manuals remain essential resources, but engagement depended on task type, modality, and layout. Despite their importance, manuals often lacked the clarity and non-visual guidance required for orientation, part differentiation, and task verification. AI tools did not address these shortcomings and sometimes worsened them with incomplete or misleading guidance. As a result, participants turned to workarounds such as YouTube tutorials, prior experience, or sighted assistance, but these approaches also had limitations that reduced their effectiveness and often prevented independent task completion. Informed by these findings, we proposed design directions for both product manufacturers and AI developers to reduce cognitive load and support more accurate, efficient, and independent completion of DIY tasks involving tangible products.

## Acknowledgments

We thank the anonymous reviewers for their insightful feedback. This work was supported by the Google Inclusion Research Award, NSF Award 2153056, NIH Award R01EY035688 and DoD award HT94252410098.

## A Appendix

## A.1 DIY tasks performed by users

| PID | Assembly | Learning to use a product | Troubleshooting |
| --- | --- | --- | --- |
| P1 | Bekvam spice rack (1d) | New air-fryer (1 wk) | Lights not turning on through the new Google assistant (last evening) |
| P2 | 2-shelf Mini Clothes organizer (1 wk) | USCCE Alarm Clock with FM, blacktooth speaker and wireless charger (2d) | Built-in blacktooth speaker of the Alarm Clock not pairing (2d) |
| P3 | Small Shoe rack (2-3d) | Touch screen refrigerator (4-5d) | Smart Thermostat beeping continuously (1 wk) |
| P4 | RASKOG Utility cart (few hours ago) | Honeywell Thermostat (1 wk) | Network connection failure on a smart TV (2-3d) |
| P5 | 2-shelf Mini-dresser (2 wk) | HUMHOLD Portable A/C (1 wk) | - |
| P6 | Letter tray (1d) | Talking microwave (3-4d) | Unable to switch the microwave mode to kitchen timer mode (last night) |
| P7 | TERTIAL work lamp (2 wk) | Portable Espresso Coffee Maker (2d) | - |
| P8 | Table-top bookshelf (2-3 wk) | A/C (3 wk) | Airfyer beeping with "SHY" displayed on the screen (four hours ago) |
| P9 | Pine crate (2d) | Slow cooker (3-4 wk) | - |
| P10 | Step Stool (1-2 wk) | Breville Control Freak Induction Cooker (1d) | - |
| P11 | Vacuum cleaner (3d) | Pet camera with treat dispenser (2-3d) | - |
| P12 | Clothes organizer (2 wk) | Printer (1 wk) | Robot Vacuum not powering up (1-2 wk) |
| P13 | Mini Kitchen Rack (5d) | Sound Bar (2-3d) | - |
| P14 | Table-top File Organizer (4d) | Slow cooker (2d) | - |
| P15 | Vaccum Cleaner (2-3 wk) | Ninja Air-fryer (3d) | - |

Table 6: Examples of DIY assembly, learning, and troubleshooting activities reported by participants that involved product manuals, along with when they last performed each task relative to the interview. A dash (–) indicates the participant did not recall performing that activity with a manual in the past two months.

## A.2 Participants' Prompts

| PID / Level | Task | User Prompt | AI Response |
| --- | --- | --- | --- |
| P7 (Beginner) AI tool : Be My AI | A1 | *You said B is a rectangular panel. I can feel one wooden panel and many rectangular meshes with little circles and rods. Can you tell me which one I should use? (After uploading the image.)* | *Yes, that's a great question. I understand what you are asking for. For part B, you should use the single wooden panel, not the mesh panels. In the diagram, part B is the only wooden panel among the parts. The mesh pieces (with circles and rods) are for the sections where you'll store items vertically, like file slots or trays, which come in later assembly steps. So for step 1, use the two large wooden side panels with handle holes (A), the flat wooden panel (B), and the small flat wooden panel (C). Set aside the meshes and the rods for later steps.* |
| P3 (Beginner) AI tool : Be My AI | A1 | *(follow up) I literally told you I don't know what A, B or C is and you keep telling me that. What the heck is A and B in Step 1?* | *Thank you for sharing your frustration — I understand you want a tactile description instead of repeating the letters. Here's how you can identify them by touch. Part A: Feel for two large panels, each one will be flat and thick, with a wide surface. You should find a rectangular oval hole (like a handle you can put your fingers through). These two identical panels are the sides. Part B: It is screw. You should make an oval frame by connecting panels A, D, and M as shown in step 1, and then below this frame, you should install E so that it can latch on to the sides of two large wooden panels frame and hang so that A, B, C and D are on E.* |
| P2 (Intermediate) AI tool : Be My AI | T2 | *Where is AL1? Explain Step 4 with details on how to locate the components.* | *The AL1 is located on ALARM MODE switch on the Sonic Bomb alarm clock.* |
| | | *(follow-up)* | |
| | | *Where is the ALARM MODE switch? Explain Step 4 again clearly. Please provide details on how to locate the components involved in Step 4.* | *AL1 is located on the ALARM MODE button on the Sonic Bomb alarm clock. The switch is near the Snooze button; you should see 'AL1' printed next near it by the snooze bar. Slide the switch toward the AL1 label to select that alarm.* |
| P5 (Intermediate) AI tool : Be My AI | A1 | *Explain the Step 1. Please mention the components involved. Don't rely on colors or labels. Give me in terms of what I can touch and feel.* | *Step 1 requires to take both the panels A, keep it vertical and slide B, C and D and screw them in place using the connector M. Part A: This is a rectangular panel, and the sheet indicates that there are 2 pieces of part A included in the packaging. It is shown at the top left corner of the packaging diagram. rectangular mesh piece. Similarly Part B, C and D are all rectangular mesh pieces that act as shelves. Pick up each one and compare it with the picture in the manual to confirm you have the correct part before assembling.* |
| | | *(follow-up)* | |
| | | *How do I distinguish between B, C, and D by touch? Please don't rely on colors or labels. Give me in terms of what I can touch and feel.* | *Sure! Here is a more specific description of parts B, C and D based on the instruction sheet: B: This is one medium-sized, flat panel, but smaller than A. C: This is a square flat mesh. It likely serves as the bottom of the organizer. D: This is another rectangular shelf. You should have two of these. Let me know if you'd like more help identifying the other parts!* |

| PID / Level | Task | User Prompt | AI Response |
|---|---|---|---|
| P1 (Expert) AI tool : Seeing AI | A2 | *Explain Step 1 again clearly. Do not use anything that requires vision. No colors, no 'you should see'. Do not repeat what is exactly present in the manual. Be very precise.*<br><br>*(follow-up using ChatGPT)*<br><br>*I am blind. Tell me exactly where to attach the phone holder and how to know if what I have done is correct or not. Additional information: I am in step 1. There are two holes on the ring, one on the inner surface and one on the outer surface. Should I insert the phone-holder mount into the inner or outer edge?* | *Step 1: Attach the phone holder with pivot mount by screwing mount into the ring light.*<br><br>*The image shows a laptop screen displaying a document which includes consists of two images. On the right most image shows phone holder with pivot mount being attached to inner edge. Hold the ring so it faces you and insert the mount into the hole on the inner surface, then tighten it until it looks straight.* |
| P4 (Expert) AI tool : ChatGPT | T1 | *I am blind. Give simple, precise steps for setting 7 p.m. on this timer dial. Use a clock-face description on the dial surface only. Remember – no colors, no 'you should see'. (After uploading the image.)* | *On the dark side of the dial you will find 7 PM marked. Turn the dial until the pointer is at that mark, then push down the pins at that point to set the timer. When the pointer looks aligned with the 7 PM marking, the time is set.* |

Table 7: Examples of participant prompts and corresponding AI responses across prompting styles. Across levels, AI responses frequently reverted to vision-dependent cues, vague spatial descriptions, and label reuse, offering limited non-visual guidance.

## A.3 Design Guidelines

Building on the two complementary directions outlined in Section 6.2, Table 8 distills the product-manual recommendations into concrete, actionable guidelines, paired with design suggestions for implementing them and the potential benefits for blind DIY-ers.

| Design Guidelines | Design Suggestions | Potential Benefits |
| --- | --- | --- |
| Use a simple, stable instruction structure | Format instructions as single-column booklets; present one action per step; avoid fold-outs and dense layouts | Supports accurate, efficient task execution; reduces the chance of skipped steps |
| Provide tactile structure and progress tracking | Add raised edge markers to indicate content type (e.g., assembly vs. troubleshooting); include tactile checklists users can mark while working | Provides a persistent record of progress, reduces cognitive load, and supports error recovery |
| Offer interactive, accessible support alongside print | Include QR codes that open accessible digital interfaces for conversational Q&A and context-sensitive guidance (e.g., "how do I attach the left bracket?") | Enables just-in-time clarification; reduces task abandonment when users get stuck |
| Anchor guidance to the physical components | Embed low-cost NFC tags in parts; tapping triggers part-specific audio cues describing orientation, assembly steps, and completion checks | Reduces the cognitive burden of mapping diagrams to components; improves accuracy and independence |

Table 8: Guidelines for redesigning product instructions to support accurate, efficient, and independent DIY task completion for blind users.

Table 9 lays out the AI-tool recommendations from Section 6.2, including layered instruction generation, grounded orientation cues, and retrieval/structured prompting, along with implementation suggestions and potential benefits.

| Design Guidelines | Design Suggestions | Potential Benefits |
| --- | --- | --- |
| Layered instruction generation | Generate instructions in layers: *macro* (main components + spatial relations), *meso* (sub-component links/constraints), *micro* (tactile cues such as grooves, textures, locking clicks); expand datasets like CheckManual with these annotations to fine-tune models [89] | Improves the operationalizability of instructions for DIY tasks by including orientation, connections, and tactile landmarks that VLM outputs often miss |
| Ground orientation in user-relative frames and identifiable feature anchors | Express orientation relative to the user and tie it to stable, identifiable features (e.g., "rotate until the flat panel faces forward and the socket points downward"); avoid label- or diagram-only references | Reduces ambiguity and improves reproducibility across users' handling styles and viewpoints |
| Use spatially informed representations to strengthen grounding when possible | Leverage 3D scene graphs and depth-enhanced encoders to improve spatial reasoning and to ground orientation cues in product images/diagrams [34] | Produces more consistent spatial guidance and reduces orientation and placement errors |
| Deliver task-relevant steps through manual segmentation and content retrieval | Use prompt chaining to segment manuals into task-based sections [76]; apply retrieval-augmented generation to surface only the relevant steps in response to a user query [81] | Limits information overload and helps users quickly access the right step at the right moment |
| Enforce structured, step-by-step guidance | Combine few-shot prompting [29] with Chain-Of-Thought [130] within a structured template (Figure 4) to produce stepwise guidance and explicit checks | Improves clarity during troubleshooting and makes corrective actions easier to follow non-visually |
| Extend beyond text to continuous, real-time assistance | Use smart glasses with real-time video analysis to monitor progress continuously, provide context-sensitive guidance, detect errors, and suggest corrections | Reduces reliance on static image captures, enables timely corrections and improves safety and efficiency in completing the tasks |

Table 9: Guidelines for AI tools to support efficient, accurate, and independent DIY task completion for blind users.